# OPEN DATA, PRIVACY, AND FAIR INFORMATION PRINCIPLES: TOWARDS A BALANCING FRAMEWORK

*Frederik Zuiderveen Borgesius, Jonathan Gray & Mireille van Eechoud†*

## ABSTRACT

Open data are held to contribute to a wide variety of social and political goals, including strengthening transparency, public participation and democratic accountability, promoting economic growth and innovation, and enabling greater public sector efficiency and cost savings. However, releasing government data that contain personal information may threaten privacy and related rights and interests. In this Article we ask how these privacy interests can be respected, without unduly hampering benefits from disclosing public sector information. We propose a balancing framework to help public authorities address this question in different contexts. The framework takes into account different levels of privacy risks for different types of data. It also separates decisions about access and re-use, and highlights a range of different disclosure routes. A circumstance catalogue lists factors that might be considered when assessing whether, under which conditions, and how a dataset can be released. While open data remains an important route for the publication of government information, we conclude that it is not the only route, and there must be clear and robust public interest arguments in order to justify the disclosure of personal information as open data.

---

† Frederik Zuiderveen Borgesius is a post-doctoral researcher at the Institute for Information Law, University of Amsterdam Law School, The Netherlands. Jonathan Gray is a Research Associate at the Digital Methods Initiative, University of Amsterdam and Director of Policy and Research at Open Knowledge. Mireille van Eechoud is professor of Information Law at the Institute for Information Law, University of Amsterdam Law School.
We thank Simon Hania, Dariusz Kloza, Stefan Kulk, Maja Lubarda, Richard Rogers, Javier Ruiz, Nico van Eijk, Ben Worthy, and Bendert Zevenbergen for participating in the *Workshop Reconciling Fair Information Principles and Open Data Policies* on February 6, 2015 at the Institute for Information Law, Amsterdam. We also thank the participants of the symposium *Open Data: Addressing Privacy, Security, and Civil Rights Challenges,* April 17, 2015, Berkeley Center for Law & Technology, in particular Cathy O'Neil and David Flaherty. The thought-provoking discussions during both events helped to shape our ideas for this Article. Furthermore, Matthijs Koot, Bendert Zevenbergen, and the editors of the Berkeley Technology Law Journal deserve our gratitude for comments on earlier versions of this Article. We also express our gratitude to the members of the advisory board for the project that led to this Article: Simon Hania, Corporate Privacy Officer at the TomTom company; Dr. Jaap-Henk Hoepman, Associate Professor of Privacy Enhancing Protocols and Privacy by Design, University of Nijmegen; Dr. Aleecia McDonald, non-residential fellow, Center for Internet & Society, Stanford University; Prof. B. Roessler, Professor of Ethics and its History, University of Amsterdam; Javier Ruiz Diaz, Policy Director, Open Rights Group; Prof. N.A.N.M. van Eijk, Professor of Media and Telecommunications Law, University of Amsterdam; Dr. Ben Worthy, lecturer in Politics at Birkbeck University of London, independent reporter for the U.K.'s IRM of the Open Government Partnership (U.K.). We thank Sarah Eskens, Rachel Wouda, and Dirk Henderickx for research assistance. All errors are the authors' own. Financial support for this project came from the Berkeley Center for Law & Technology and Microsoft.

## TABLE OF CONTENTS

## I.    INTRODUCTION

Open government data refers to data released by public sector bodies, in a manner that is legally and technically re-usable. The *G8 Open Data Charter* states "free access to, and subsequent re-use of, open data are of significant value to society and the economy."[1] Open data are commonly held by its advocates to mean data that "can be freely used, modified, and shared by anyone for any purpose."[2] However, releasing public sector datasets that include personal information, or data that can be re-identified, may threaten privacy and related rights.

In this Article, we examine the tension between public sector open data policy and the Fair Information Principles (FIPs). The FIPs lie at the core of most data privacy laws around the world, including those in the European Union and the United States. The FIPs give guidelines to balance privacy-related interests and other interests, such as those of business and the public sector. The Article focuses on the following question: from the perspective of the Fair Information Principles, how can privacy and related interests be respected, without unduly hampering benefits from disclosing public sector information?

We rely mostly on desk research, using the usual sources for legal scholarship, such as legislation, soft law, policy documents, and literature.

---

    1. G8 Open Data Charter (2013) https://www.gov.uk/government/publications/open-data-charter/g8-open-data-charter-and-technical-annex.
    2. Open Definition, http://opendefinition.org (last visited May 1, 2015). Open Knowledge's first open definition dates from 2005. *Open Knowledge Definition 1.0*, Open Definition, http://opendefinition.org/od/1.0 (last visited May 1, 2015).

We use descriptive and analytical legal research to determine the main legal tensions between open data policy and the FIPs. Parts of the Article are more normative: we give recommendations to strike a balance that respects privacy and related interests, and does not unduly hamper the benefits of open data.

We enriched our research results with insights from a workshop, where we tested hypotheses and discussed the promises and pitfalls of privacy and open data. Conference participants came from academia, industry, civil society organizations, and data protection authorities, and were all working on issues in open data and privacy.[3] Discussions during the *Open Data: Addressing Privacy, Security, and Civil Rights Challenges* symposium held by the Berkeley Center for Law & Technology also provided valuable insights.[4]

Furthermore, we conducted an empirical study into concerns that various stakeholders, in civil society, the public sector, research, and business, express about the interactions between privacy and open data. The study draws on document collections and digital traces from the web to map the debates about privacy and open data. The empirical study follows the "digital methods" approach, pioneered by Richard Rogers and his colleagues at the Digital Methods Initiative.[5]

While each national legal system has its own traditions and characteristics, this Article focuses on common problems that arise in many jurisdictions. After all, as the Open Government Partnership (OGP) testified, governments around the world create open data policies and must cope with privacy concerns.[6] Hence, we do not examine what sets jurisdictions apart, but instead discuss shared problems. For instance, we do not address specific requirements that follow from the First

---

3. Workshop, Reconciling Fair Information Principles and Open Data Policies, Institute for Information Law, Amsterdam, Netherlands (Feb. 6, 2015).

4. *See Addressing Privacy, Security, and Civil Rights Challenges—19th Annual BCLT/BTLJ Symposium*, BERKELEY LAW (Apr. 17, 2015), https://www.law.berkeley.edu/centers/bclt/past-events/april-2015-the-19th-annual-bcltbtlj-symposium-open-data-addressing-privacy-security-and-civil-rights-challenges/program.

5. *See generally* RICHARD ROGERS, DIGITAL METHODS (2013).

6. The OGP is an international platform for reform, to make "governments more open, accountable, and responsive to citizens." Participating states submit action plans in which they make commitments, inter alia on datasets to be made available as open data. Compliance and progress mechanisms are in place. Membership has grown to 65 countries in the five years since the OGP's inception. *See* OPEN GOVERNMENT PARTNERSHIP, http://www.opengovpartnership.org (last visited May 1, 2015).

Amendment in the United States,[7] or from the fundamental right to data protection in the European Union.[8] Therefore, the Article's recommendations come with a caveat: they cannot be directly implemented in national legal systems.

The Article is structured as follows. Part II describes open data goals and privacy problems regarding open data. We clustered the objectives associated with open data into three categories: (1) innovation and economic growth, (2) political accountability and democratic participation, and (3) public sector efficiency. We identified three kinds of concerns about releasing personal information as open data: (1) the chilling effects on people interacting with the public sector, (2) a lack of individual control over personal information, and (3) the use of open data for social sorting or discriminatory practices.

Part III discusses rules regarding access to information held by public sector. Freedom of information laws provide inspiration on how to strike a balance between privacy and transparency in the open data context.

Part IV discusses the governance of personal information, focusing on the Fair Information Principles (FIPs). In this section we also discuss the main challenges in reconciling open data policy and the FIPs. From a FIPs perspective, the main problem with open data is that unrestricted re-use of personal data breaches the purpose specification principle. But we argue that there are possible compromise measures to balance privacy and open data interests.

We propose a balancing framework to accommodate privacy concerns and open data goals. Part V outlines the first element of the balancing framework, and distinguishes four data categories with different levels of privacy risks: (A) raw personal data, (B) pseudonymized data, (C) anonymized data, and (D) non-personal data. Different modes of access and re-use control are the second element of the balancing framework. In many cases, disclosing data with access or re-use restrictions, rather than as fully open data, strikes a balance between open data goals and privacy (Part VI). As a third element of the balancing framework we provide a circumstance catalogue, a list of circumstances to consider when deciding

---

7. *See* Daniel J. Solove, *Access and Aggregation: Public Records, Privacy and the Constitution*, 86 MINN. L. REV. 1137, 1201 (2002).

8. For more on EU data protection law and public sector information re-use policy, see Cristina Dos Santos et al., *On Privacy and Personal Data Protection*, 6 MASARYK U. J.L. & TECH. 337 (2012), https://journals.muni.cz/mujlt/article/view/2613/2177; *see also* Mireille van Eechoud et al., *LAPSI Position Paper on Access to Data*, LAPSI (Dec. 12, 2014), http://dare.uva.nl/document/2/162858.

whether or not a dataset should be disclosed, and under which conditions (Part VII).

Part VIII concludes that releasing personal information as fully open data is generally not appropriate. But sometimes a compromise can be found by disclosing data with access or re-use restrictions.

## II. OPEN DATA AND PRIVACY

Open data are held to contribute to a wide variety of social and political goals. However, releasing data as open data may threaten privacy, for instance, if the open data contain personal information. Below we describe open data goals and privacy problems regarding open data. We clustered the objectives associated with open data into three categories: (1) innovation and economic growth, (2) political accountability and democratic participation, and (3) public sector efficiency. We also clustered privacy concerns in the area of open data into three categories: (1) the chilling effects on people interacting with the public sector, (2) a lack of individual control over personal information, and (3) the use of open data for social sorting or discriminatory practices.

### A. OPEN DATA INTERESTS

Definitions of open data from technologists and civil society actors focus on enabling redistribution and re-use, and on limiting legal and technical barriers to re-use. For example, the summary of the "Open Definition" from Open Knowledge reads: "Open means anyone can freely access, use, modify, and share for any purpose (subject, at most, to requirements that preserve provenance and openness)."[9] The full definition stipulates conditions that include legal openness, bulk downloadability, and machine-readability.[10] Similar definitions are used in the *8 Principles of Open Government Data*,[11] the Sunlight Foundation's *Ten Principles for Opening Up Government Information*,[12] and the World Wide Web Consortium's *Five Stars of Linked Open Data*.[13]

---

9. OPEN DEFINITION, *supra* note 2.
10. *Id.*
11. OPEN DATA WORKING GROUP, *The 8 Principles of Open Government Data*, OPENGOVDATA.ORG (Dec. 8, 2007), http://opengovdata.org.
12. *Ten Principles for Opening Up Government Information*, SUNLIGHT FOUNDATION (Aug. 11, 2010), http://sunlightfoundation.com/policy/documents/ten-open-data-principles.
13. Tim Berners-Lee, *Linked Data*, W3.ORG (June 18, 2009), http://www.w3.org/DesignIssues/LinkedData.html.

Technical obstacles for re-using data include non-machine readable formats, proprietary formats, technological protection mechanisms,[14] and Digital Rights Management software. Legal restrictions on re-use include intellectual property rights, such as copyright and database rights.[15] When open data advocates say that "anyone can freely access, use, modify, and share [data] for any purpose,"[16] they are often referring to removing these specific kinds of legal and technical restrictions.

This conception of open data that focuses on limiting legal and technical restrictions for re-use has carried into public policy. Over the past decade, open data developed from being a niche idea at the margins of open source software, scientific research and hacker communities, into an idea with traction among public policymakers.[17] For example, the 2013 *G8 Open Data Charter* mentions that open data should be "machine readable," available in bulk, available in formats for which the specification is "available to anyone for free," and under open licenses such that "no restrictions or charges are placed on the re-use of the information for non-commercial or commercial purposes."[18] A similar focus on removing technical restrictions to re-use can be found in open data guidelines of the Organisation for Economic Co-operation and Development,[19] the U.K. government,[20] and U.S. President Barack Obama.[21]

---

14. Technological protection mechanisms (TPMs) and digital rights management information are protected against circumvention and interference in their own right, separate from, e.g., copyright in the underlying work (database, software or other works). *See* Berne Convention for the Protection of Literary and Artistic Works, art. 11, 12, *as amended* Sept. 28, 1979, S. TREATY DOC. No. 99-27; WIPO Copyright Treaty, Dec. 20, 1996, S. TREATY DOC. No. 105-17.

15. There is controversy about the role of intellectual property rights in implementing public sector open data, but this controversy is beyond the scope of this Article.

16. OPEN DEFINITION, *supra* note 2.

17. Jonathan Gray, Towards a Genealogy of Open Data (Sept. 3, 2014) (Conference Paper given at the General Conference of the European Consortium for Political Research in Glasgow, Scotland), http://dx.doi.org/10.2139/ssrn.2605828.

18. G8 OPEN DATA CHARTER, *supra* note 1.

19. *See* Barbara Ubaldi, *Open Government Data: Towards Empirical Analysis of Open Government Data Initiatives* (Organisation for Economic Co-operation & Development, Working Paper on Public Governance No. 22, 2013), http://dx.doi.org/10.1787/5k46bj4f03s7-en; *see also* Organisation for Economic Co-operation & Development [OECD], *OECD Recommendation of the Council for Enhanced Access and More Effective Use of Public Sector Information*, OECD Doc. C(2008)36 (2008), https://www.oecd.org/sti/44384673.pdf [hereinafter *OECD Recommendation*].

20. *Public Data Principles*, DATA.GOV.UK (Apr. 10, 2012), http://data.gov.uk/library/public-data-principles.

21. *See* Exec. Order No. 13,642, Making Open and Machine Readable the New Default for Government Information, 78 Fed. Reg. 28111 (May 9, 2013), https://www

Open data are held to contribute to a wide variety of social and political goals.[22] For ease of discussion in this Article, we have clustered the many objectives associated with open data into the following three areas: (1) innovation and economic growth, (2) political accountability and democratic participation, and (3) public sector efficiency. First we look at fostering innovation and economic growth.

### 1. Innovation and Economic Growth

Most official open data initiatives highlight the potential of enabling the re-use of public sector information to create new businesses and innovative services and products. Open data policies are increasingly becoming the preferred route to unlock the value of public sector information. This is evident from the European Commission's Guidelines on the Public Sector Information Directive.[23] President Obama's 2013 executive order, which aims to make Open and Machine Readable the New Default for Government Information, views (federal) government information as a national asset and recognizes the importance of enabling widespread re-use for "economic growth and job creation."[24] President Obama's 2013 executive order on Open Data Policy adds: "making information resources accessible, discoverable, and usable by the public can help fuel entrepreneurship, innovation, and scientific discovery."[25] Similarly, the *G8 Open Data Charter* claims open data are "a catalyst for innovation in the private sector, supporting the creation of new markets, businesses, and jobs."[26] The World Bank also recognizes this potential of open data.[27]

---

.whitehouse.gov/the-press-office/2013/05/09/executive-order-making-open-and-machine-readable-new-default-government- [*hereinafter* Exec. Order, Open and Machine Readable].

22. *See, e.g.*, Gray, *supra* note 17.
23. Commission Notice: Guidelines on Recommended Standard Licenses, Datasets and Charging for the Re-Use of Documents, 2014 O.J. (C 240) 1.
24. Exec. Order, Open and Machine Readable, *supra* note 21. The Order is one of several that follow up on open government policy announced by the White House in January 2009. Memorandum for the Heads of Executive Departments and Agencies on Transparency and Open Government, 74 Fed. Reg. 15 (Jan. 21, 2009).
25. OFFICE OF MGMT. & BUDGET, EXEC. OFFICE OF THE PRESIDENT, OMB MEMORANDUM M-13-13, OPEN DATA POLICY—MANAGING INFORMATION AS AN ASSET (2013), http://www.whitehouse.gov/sites/default/files/omb/memoranda/2013/m-13-13.pdf [hereinafter OMB MEMORANDUM M-13-13, OPEN DATA POLICY].
26. *See* G8 OPEN DATA CHARTER, *supra* note 1. It was signed by G8 leaders on June 18, 2013 to promote transparency, innovation, and accountability.
27. WORLD BANK, OPEN DATA FOR ECONOMIC GROWTH 5 (2014).

Information services built on public sector data are diverse. Financial services providers use official statistics as input.[28] Companies in the meteorological sector use weather data to provide highly specialized services, e.g., forecasts for off-shore oil industries.[29] Planning permissions, zoning data and housing data are combined with other sources to produce advice for customers such as real estate developers.[30] Postal codes are widely used as identifiers.[31] School and health inspection data serve as input for apps that help inform parents or patient choice.[32] Public transport timetable data when combined with geolocation data enable real-time and customized travel advice.[33] There are many other kinds of commercial exploitation of open data, often involving the combination of data from different public and private sources to deliver information products or services. The emphasis on economic benefits of re-using data held by public sector bodies predates open data policies. For example, in 1989, the E.U. sought to stimulate commercial exploitation of public sector data by the private sector.[34] The E.U. Public Sector Information Directive of 2003 also focused on public sector information as raw material for creating services and products.[35] The Directive obliged a wide range of public sector bodies to allow commercial and non-commercial re-use of their information assets, but not necessarily as open data.[36] Under the directive, conditions may be imposed, costs charged, and data may be made available in non-structured form. The U.S. Office of Management & Budget also recognized federal information as a "commodity in the marketplace."[37]

---

28. For examples of government information re-use, see MARTIN FORNEFELD ET AL., MICUS, REPORT FOR THE EUROPEAN COMMISSION, ASSESSMENT OF THE RE-USE OF PUBLIC SECTOR INFORMATION (PSI) (2009); MAKX DEKKERS ET AL., MEASURING EUROPEAN PUBLIC SECTOR INFORMATION RESOURCES (MEPSIR), REPORT FOR THE EUROPEAN COMMISSION, FINAL REPORT OF STUDY ON EXPLOITATION OF PUBLIC SECTOR INFORMATION 37 (2006).

29. For example, consider the private company MeteoGroup. *See Marine*, METEOGROUP, http://www.meteogroup.com/en/gb/sectors/marine.html (last visited May 15, 2015).

30. For example, in Europe, the company Landmark provides such services and took the city of Amsterdam to court for the price it charged for re-use of city data. *See* ABRvS 20 april 2009, AB 2009, 546 m.nt. JJB (B&W Amsterdam/Landmark) (Neth.).

31. For this reason the G8 Open Data Charter lists postal codes as "high value" data, to be made available with priority. G8 OPEN DATA CHARTER, *supra* note 1.

32. U.S. DEP'T OF COMMERCE, FOSTERING INNOVATION, CREATING JOBS, DRIVING BETTER DECISIONS: THE VALUE OF GOVERNMENT DATA (2014); G8 OPEN DATA CHARTER, *supra* note 1; MCKINSEY & CO., OPEN DATA: UNLOCKING INNOVATION AND PERFORMANCE WITH LIQUID INFORMATION 11 (2013).

33. *See* MCKINSEY & CO., *supra* note 32, at 6.

34. Directorate Gen. for Telecomm., Info. Indus. & Innovation, Comm'n of the European Cmtys., *Guidelines for Improving the Synergy Between the Public and Private Sectors in the Information Market* (1989).

35. Directive 2003/98/EC, of the European Parliament and of the Council of 17 November 2003 on the Re-use of Public Sector Information, 2003 O.J. (L 345) 90 (revised by Directive 2013/37/EC, 2013 O.J. (L 175) 1).

Many studies have been commissioned to assess the value of public sector information; these studies suggest impressive figures, but range widely.[38] For example, the U.S. Department of Commerce looked at the size of private sector revenues from "government data-intensive business activities" for the United States and arrived at a crude estimate in the range of 24 to 221 billion USD per year.[39] And a 2000 study for the European Commission estimated that for the then 15 E.U. member states, the part of the combined national income attributable to industries and activities built on exploiting public sector information ranged between €28 billion and €134 billion. Some have judged these estimates as far too optimistic.[40] Generally, researchers recognize there is a lack of hard data on which to base estimates.[41] Nevertheless, policymakers see fostering innovation and economic growth as an important goal of open data.

---

36. The *obligation* to allow re-use was introduced in the 2013 revision. Directive 2013/37/EC, 2013 O.J. (L 175) 1. Member states must implement the revised directive by July 2015. The Directive builds on public access regimes in member states; it does not regulate access directly.

37. Office of Mgmt. & Budget, Exec. Office of the President, OMB Circular No. A-130 Revised, MANAGEMENT OF FEDERAL INFORMATION RESOURCES (1998). First issued in 1985, the Circular fostered (among many things) a larger role for the private sector in dissemination of government information and creating added-value (electronic) services. With subsequent revisions (1993–1996) under the Clinton administration the focus moved to release of electronic information by federal agencies directly to the public. For an overview of early policy development, see U.S. OFFICE OF TECHNOLOGY ASSESSMENT, OTA-C IT-396, INFORMING THE NATION: FEDERAL INFORMATION DISSEMINATION IN AN ELECTRONIC AGE (1988) [*hereinafter* FEDERAL INFORMATION DISSEMINATION IN AN ELECTRONIC AGE].

38. For recent examples of studies on the economic value of public sector information at the E.U. level, see MARC DE VRIES ET AL., PRICING OF PUBLIC SECTOR INFORMATION. MODELS OF SUPPLY AND CHARGING FOR PUBLIC SECTOR INFORMATION, FINAL REPORT (2011); MARC DE VRIES ET AL., REPORT FOR EUROPEAN COMMISSION, PRICING OF PUBLIC SECTOR INFORMATION STUDY (POPSIS) (2011). For recent examples of studies about the value of open data and public sector information at the national level, see U.S. DEP'T OF COMMERCE, *supra* note 32; DELOITTE, MARKET ASSESSMENT OF PUBLIC SECTOR INFORMATION, STUDY FOR U.K. DEPARTMENT FOR BUSINESS, INNOVATION, & SKILLS (2013); U.K. OFFICE OF FAIR TRADING, OFT861, THE COMMERCIAL USE OF PUBLIC INFORMATION (2006). For examples of subnational level studies, see Jens PREISCHE, DIGITALES GOLD: NUTZEN UND WERTSCHÖPFUNG DURCH OPEN DATA FÜR BERLIN (2014); Gregor Eibl & Brigitte Lutz, *Money for Nothing—Data for Free: Hard Facts About the Economic Power of Open Government Data*, in CEDEM13: CONFERENCE FOR E-DEMOCRACY AND OPEN GOVERNMENT 289 (Peter Parycek & Noelle Edelmann eds., 2d ed. 2013).

39. U.S. DEP'T OF COMMERCE, *supra* note 32.

### 2. Political Accountability and Democratic Participation

A second goal pursued through open data policy is fostering political accountability and democratic participation. Current proactive disclosure policies cover a broad range of information: from basic information about a public authority's responsibility, organization, and procedures, to granular data about public spending and subsidies awarded.[42]

In the open data context, statements about the perceived benefits of open data for democracy are frequent. The *G8 Open Data Charter* mentions good governance and anti-corruption,[43] and argues that more public data on the use of natural resources and distribution of revenues, on land management, and on development spending would promote accountability and good governance.[44] The World Bank makes a similar case, arguing that open data "supports democratic societies" and "encourages greater citizen participation in government affairs."[45] The French government's open data policy is driven by the idea that "opening and sharing data is the way for modern government to organize itself so that it is accountable, opens dialogue and trusts the collective intelligence of its citizens."[46] The Obama administration posits that making information available proactively online in open formats increases accountability and promotes informed participation by the public.[47] A basic consideration of policy for the management of U.S. federal information is that public disclosure of government information is essential to the operation of a democracy.[48] Similarly, the E.U. Public Sector Information Directive says that publishing documents held by the public sector "is a fundamental instrument for extending the right to knowledge, which is a basic principle of democracy."[49]

---

40. Robbin te Velde, *Public Sector Information: Why Bother?*, *in* THE SOCIO-ECONOMIC EFFECTS OF PUBLIC SECTOR INFORMATION ON DIGITAL NETWORKS: TOWARD A BETTER UNDERSTANDING OF DIFFERENT ACCESS AND REUSE POLICIES: WORKSHOP SUMMARY 25, 25–28 (P. Uhlir ed., 2009).

41. *See* Mireille van Eechoud, *Calculating and Monitoring the Benefits of Public Sector Information Re-use*, *in* ZUGANG UND VERWERTUNG ÖFFENTLICHER INFORMATIONEN (Thomas Dreier et al. eds., forthcoming 2015).

42. *See, e.g.*, *Cabinet Office Organogram*, DATA.GOV.UK, http://data.gov.uk/organogram/cabinet-office; *Senior Officials "High Earners" Salaries*, DATA.GOV.UK, http://data.gov.uk/dataset/uk-civil-service-high-earners; *Where Does Europe's Money Go? A Guide to EU Budget Data Sources*, OPEN KNOWLEDGE BLOG (July 2, 2015), http://blog.okfn.org/2015/07/02/where-does-europes-money-go.

43. G8 OPEN DATA CHARTER, *supra* note 1, ¶¶ 4–5.

44. *Id.*

45. Open Data Toolkit, WORLD BANK, http://opendatatoolkit.worldbank.org/en/starting.html.

46. This language is translated from "L'ouverture et le partage des données, c'est la manière, pour un Etat moderne, de s'organiser afin de rendre des comptes, d'ouvrir le dialogue, et de faire confiance à l'intelligence collective des citoyens." SÉCRETARIAT GÉNÉRAL POUR LA MODERNISATION DE LA FONCTION PUBLIQUE, VADE-MECUM: SUR L'OUVERTURE ET LE PARTAGE DES DONNÉES PUBLIQUES 5 (2013).

The idea of open government is tied to the ideal of transparency of governments' decisions and activities. Transparency is widely regarded as a precondition for the effective exercise of political rights and freedoms, and for ensuring accountable public authorities.[50] Access to information is a key aspect of democratic institutions that are based on representation, delegation, and accountability. Assessing, debating, and sanctioning public sector behavior requires accurate information.[51] In sum, the proactive disclosure of government data to the public for the purposes of political transparency, accountability and participation is becoming a central tenet in democratic governance.

---

47. OFFICE OF MGMT. & BUDGET, EXEC. OFFICE OF THE PRESIDENT, OMB MEMORANDUM M-10-06, OPEN GOVERNMENT DIRECTIVE (2009), https://www.whitehouse.gov/sites/default/files/omb/assets/memoranda_2010/m10-06.pdf.

48. *See* OFFICE OF MGMT. & BUDGET, EXEC. OFFICE OF THE PRESIDENT, OMB CIRCULAR NO. A-130 REVISED, TRANSMITTAL 2, MANAGEMENT OF FEDERAL INFORMATION RESOURCES (1994) (older revision of the Circular); OFFICE OF MGMT. & BUDGET, EXEC. OFFICE OF THE PRESIDENT, OMB CIRCULAR NO. A-130 REVISED, TRANSMITTAL 4, MANAGEMENT OF FEDERAL INFORMATION RESOURCES (2000) (current version of the Circular). The Circular has a residual role: it does not affect disclosure duties or rights to information under FOIA.

49. Recital 16 of the PSI Directive states:

> Making public all generally available documents held by the public sector—concerning not only the political process but also the legal and administrative process—is a fundamental instrument for extending the right to knowledge, which is a basic principle of democracy. This objective is applicable to institutions at every level, be it local, national or international.

Directive 2003/98/EC, *supra* note 35.

50. For an expanded discussion of transparency, see CHRISTOPHER HOOD, & DAVID HEALD, TRANSPARENCY: THE KEY TO BETTER GOVERNANCE? (2006); MARK BOVENS ET AL., THE OXFORD HANDBOOK OF PUBLIC ACCOUNTABILITY (2014).

51. Like transparency, accountability is a multifaceted concept. For a discussion of dimensions in relation to democracy, see Gijs Jan Brandsma & Thomas Schillemans, *The Accountability Cube: Measuring Accountability*, 23 J. PUB. ADMIN. RES. THEORY 953 (2013).

### 3. *Public Sector Efficiency and Service Delivery*

A third set of pro open data arguments focuses on efficiency: open data should help to save resources and improve public services. For instance, the European Commission says open data will improve health services and traffic management, and help tackle environmental challenges, for instance through monitoring energy consumption.[52]

At the national level, an increasingly popular strategy is to publish performance data of publicly funded organizations.[53] Disclosing inspection and other data is alleged to improve performance of recipients of tax monies, like schools (test scores) and hospitals (deaths, waiting times).[54] Citizens in their capacity as customers are presumed to make better-informed choices when provided with such performance data.[55] Other initiatives serve to improve compliance and to assist in better policymaking or prioritizing enforcement, for instance in the area of food safety standards or building safety.[56] Some open government data initiatives propose a more active role for the public: as an army of armchair auditors who can help identify possible savings.[57]

Furthermore, open data are expected to help public sector bodies carry out their tasks. Many users of open data portals are from the public sector.[58] Efficiency gains made when more transparency about information resources leads to less duplication of information collection, and hence more shared use of resources, are said to improve public sector services.[59] Furthermore, public sector bodies are expected to improve their services when they have more information at their disposal.[60] Efficient use of information resources is not a new concern of governments. For several decades information management policies have been argued to increase government efficiency.[61]

---

52. *Communication from the Commission to the European Parliament et al. on Open Data: An Engine for Innovation, Growth and Transparent Governance*, at 3, COM (2011) 882 final (Dec. 12, 2011), http://eur-lex.europa.eu/LexUriServ/LexUriServ.do?uri= COM:2011:0882:FIN:EN:PDF.

53. Mireille van Eechoud, Inaugural Lecture at the Institute for Information Law at University of Amsterdam: De Lokroep van Open Data 9 (May 23, 2014), http://www.ivir.nl/publicaties/download/1407.

54. *Id.* at 9.

55. MCKINSEY & CO., *supra* note 32, at 83–85.

56. *See, e.g.*, Michael Flowers, *Beyond Open Data: The Data-Driven City*, in BEYOND TRANSPARENCY 185 (Brett Goldstein & Lauren Dyson eds., 2013),

57. *See* BEN WORTHY, DAVID CAMERON'S TRANSPARENCY REVOLUTION? 9 (2013), http://doi.org/10.2139/ssrn.2361428.

58. *See* WORLD BANK, *supra* note 27.

59. McKinsey, *supra* note 32, at 57–58 (making this case for the energy sector).

60. *See* Alan Feuer, *The Mayor's Geek Squad*, N.Y. TIMES (Mar. 23, 2013), http://www.nytimes.com/2013/03/24/nyregion/mayor-bloombergs-geek-squad.html (discussing the advantages of combining existing data to yield useful information for, e.g., disaster relief efforts or environmental pollution).

In an empirical mapping study, we found that different arguments for open data obtain varying levels of attention amongst different actors in different forms of digital media.[62] For example, in English language mainstream media outlets arguments and examples about the economic growth and technological innovation potential of open data received more attention than those related to public participation or democratic accountability. On social media platforms such as Twitter, distinct groups of actors were interested in different sets of topics around open data such that, for example, some were interested in startups and smart cities, and others were interested in transparency and open government.[63]

In sum, open data policies serve diverse interests. For the purposes of this Article, these can be clustered into: (1) innovation and economic growth, (2) political accountability and democratic participation, and (3) public sector efficiency.

B.   PRIVACY INTERESTS

At the global level, the right to privacy is protected under, for instance, the United Nations Declaration of Human Rights[64] and the International Covenant on Civil and Political Rights.[65] In the United States, the Fourth Amendment and other laws protect privacy.[66] In Europe, the European Convention on Human Rights,[67] the European Union Charter of

---

61. FEDERAL INFORMATION DISSEMINATION IN AN ELECTRONIC AGE, *supra* note 37.
62. Jonathan Gray et al., Mapping the Politics of Open Data on Digital Media (in preparation) (on file with authors).
63. *Id.*
64. Universal Declaration of Human Rights, art. 12, G.A. Res. 217A (III), U.N. Doc. A/810, at 71 (1948).
65. International Covenant on Civil and Political Rights, art. 17, Dec. 16, 1966, S. Treaty Doc. No. 95-20, 6 I.L.M. 368 (1967), 999 U.N.T.S. 171.
66. *See generally* WILLIAM CUDDIHY, THE FOURTH AMENDMENT: ORIGINS AND ORIGINAL MEANING 602–1791 (2009); DANIEL SOLOVE & PAUL SCHWARTZ, INFORMATION PRIVACY LAW 260–335 (5th ed., 2014).
67. Convention for the Protection of Human Rights and Fundamental Freedoms, art. 8, Nov. 4, 1950, 213 U.N.T.S. 222 [hereinafter ECHR] (also referred to as the European Convention on Human Rights).

Fundamental Rights,[68] national constitutions, and other laws protect privacy.[69]

Public sector bodies hold an enormous amount of personal information, and this amount will likely grow. For instance, so-called "smart cities" may provide the public sector information about people such as up-to-date location data of cars, and detailed electricity metering data.[70] And, as public sector bodies offer more services online, they will obtain even more information about people.[71] Sometimes citizens volunteer personal information, for example when they use public services. But public authorities can also collect information through third parties, like educational and health care institutions.[72] And authorities can compel citizens to provide personal information. This element of force heightens privacy concerns.

We distinguish three broad categories of privacy concerns regarding open data: (1) the chilling effects on people in their interaction with the public sector, (2) a lack of individual control over personal information, and (3) the use of open data as input for social sorting and discriminatory practices.[73]

---

68. Charter of Fundamental Rights of the European Union of the European Parliament, arts. 7–8, 2010 O.J. (C 83) 2, 1 [hereinafter E.U. Charter of Fundamental Rights].

69. *See, e.g.*, Grondwet voor het Koninkrijk der Nederlanden [Constitution of the Kingdom of the Netherlands], art. 10. Furthermore, each E.U. member state has a national data protection act implementing the European Parliament's directive "on the protection of individuals with regard to the processing of personal data and on the free movement of such data." Council Directive 95/46/EC, art. 28, 1995 O.J. (L 281).

70. A smart city has been defined "as one that has digital technology embedded across all city functions." *Definitions and Overviews*, SMART CITIES COUNCIL, http://smartcitiescouncil.com/smart-cities-information-center/definitions-and-overviews (last visited May 1, 2015); *see also* Robert G. Hollands, *Will the Real Smart City Please Stand Up? Intelligent, Progressive or Entrepreneurial?*, 12 CITY 303 (2008).

71. Teresa Scassa, *Privacy and Open Government*, 6 FUTURE INTERNET 397, 397–98 (2014).

72. *See* Solove, *Access and Aggregation*, *supra* note 7, at 1142–50 (contains an overview of federal, state, and local record collection in the United States).

73. FREDERIK ZUIDERVEEN BORGESIUS, IMPROVING PRIVACY PROTECTION IN THE AREA OF BEHAVIOURAL TARGETING 53–63 (2015). The three categories are based on that study, which does not concern open data. In this Article, we adapt the categories to the open data context.

### 1. *Chilling Effects*

First, a chilling effect can occur if people interacting with public bodies fear that their information will be stored, or will be made public.[74] For example, people might be less inclined to contact public sector agencies if they doubt that their personal data will remain confidential.[75]

People often provide personal information when engaging with public sector bodies. Public sector bodies often require information, for example, when people apply for a planning permission or business license, attempt to comply with health and safety standards, or submit tax claims or grant applications. The collection, use and exchange of personal information are part of the normal fabric of public sector activity. Many public services cannot be delivered without these activities.

People might refrain from contacting the public sector if they fear their personal information will not be kept confidential. Especially people with questions about diseases, pregnancies, drugs, financial troubles, or suicidal thoughts might refrain from asking for help. Jeff Jonas and Jim Harper illustrate the importance of communicating with the public sector without disclosing too much personal information with an example regarding a migrant.[76] Say Alice is a migrant who thinks her residence permit contains errors. If she thinks that visiting the immigration website will bring her to the attention of immigration law enforcement, she might forego looking for information. "If she cannot communicate this information anonymously, she almost certainly will not ask questions or volunteer information, denying herself help she might deserve while denying policymakers relevant information."[77] If Alice thought her data would be disclosed to others in and outside government, such a chilling effect might be greater.

By itself the chilling effect already harms the individual who refrains from an activity she might otherwise engage in. But if somebody does not seek help because of a chilling effect, for instance if someone does not seek information regarding a disease, he or she may also experience more tangible harms. People foregoing treatment of infectious diseases could harm society as a whole.

---

74. *See* KIERON O'HARA, TRANSPARENT GOVERNMENT, NOT TRANSPARENT CITIZENS: A REPORT ON PRIVACY AND TRANSPARENCY FOR THE CABINET OFFICE 24 (2011), http://www.gov.uk/government/uploads/system/uploads/attachment_data/file/61279/transparency-and-privacy-review-annex-a.pdf.

75. Jeff Jonas & Jim Harper, *Open Government: The Privacy Imperative*, in OPEN GOVERNMENT: TRANSPARENCY, COLLABORATION, AND PARTICIPATION IN PRACTICE 315, 322–23 (Daniel Lathrop & Laurel R. T. Ruma eds., 2010).

76. *Id.*

77. *Id.* at 317.

Uncertainty about what happens with one's personal information can ultimately adversely impact the quality of public services. As Teresa Scassa notes, with open data "there is a risk not only to individual privacy, but also to the relationship of trust that is meant to exist between citizens and their government."[78] Government statistics offices have realized for a long time that confidentiality of census answers is important—otherwise people might not give honest answers anymore. Trust in public authorities could diminish if people do not believe that their personal data will remain confidential.[79] In sum, open data policy could lead to a chilling effect on people communicating with the public sector, which is a privacy problem.

### 2. Lack of Control over Personal Information

A second privacy concern is that people lack control over their personal information if that information is released as open data. Publicly releasing personal information as open data can be especially troublesome because open data policy in its most liberal form implies that unlimited numbers of re-users can use the data for any purpose.

Many privacy definitions focus on individual control over personal information. For instance, Alan Westin defined privacy in 1967 as "the claim of individuals, groups, or institutions to determine when, how, and to what extent information about them is communicated to others."[80] Many scholars use similar privacy definitions.[81] The privacy as control perspective is apparent in legal practice. For instance, the U.S. Supreme Court has described privacy as "the individual's control of information concerning his or her person."[82] The German Supreme Court says a person has, in principle, the right "to determine for himself whether his personal data should be divulged or utilized."[83] Privacy as control has deeply influenced the Fair Information Principles.[84] The privacy as control perspective does not capture all the subtleties of privacy. Nevertheless, a loss of individual control over personal information is widely seen as a privacy problem.[85]

---

78. Scassa, *supra* note 71, at 408.
79. *See, e.g.*, U.S. GOV'T ACCOUNTABILITY OFFICE, GAO-01-126SP, RECORD LINKAGE AND PRIVACY: ISSUES IN CREATING NEW FEDERAL RESEARCH AND STATISTICAL INFORMATION 18 (2001).
80. ALAN F. WESTIN, PRIVACY AND FREEDOM 7 (reprint 1970) (1967).
81. *See* Charles Fried, *Privacy*, 77 YALE L.J. 475, 482 (1968) (discussing that privacy "is not simply an absence of information about us in the minds of others; rather it is the *control* we have over information about ourselves."). *See also* A.R. MILLER, THE ASSAULT ON PRIVACY: COMPUTERS, DATA BANKS, AND DOSSIERS 25 (1971) (describing privacy as "the individual's ability to control the circulation of information relating to him").
82. U.S. Dep't of Justice v. Reporters Comm. for Freedom of the Press, 489 U.S. 749, 763 (1988).

A lack of individual control over personal information can lead to subjective and objective privacy harm. Objective harm is, in Ryan Calo's words, "the unanticipated or coerced use of information concerning a person against that person."[86] The Eightmaps website provides an example of objective harm resulting from data released by the public sector.[87] Proposition 8 was a 2008 proposal to amend the California constitution with a referendum to ban gay marriage.[88] California law requires that campaign donations be published.[89] An anonymous website publisher took information regarding donors who supported Proposition 8, and overlaid that information on Google Maps.[90] The map showed information such as the donor's name, approximate location, and the amount donated. Some of the donors received death threats, or were the victim of boycotts.[91] The dissemination of correct information can already produce objective harms, but the potential of harm arising from the public release of inaccurate or false data is at least as large.

---

    83. Bundesverfassungsgericht [BVerfG] [Federal Constitutional Court] Mar. 25, 1982, BGBL. I 369, 1982 (Ger.), *translated in* E.H. Riedel, *New Bearings in German Data Protection*, 5 HUM. RTS. L.J. 94, 101 (1984).
    84. *See, e.g.*, COLIN J. BENNETT, REGULATING PRIVACY: DATA PROTECTION AND PUBLIC POLICY IN EUROPE AND THE UNITED STATES 14 (1992). *See infra* Part IV.
    85. *See, e.g.*, Fahriye Seda Gürses, Multilateral Privacy Requirements Analysis in Online Social Networks (May 2010) (Ph.D. thesis, University of Leuven); HELEN NISSENBAUM, PRIVACY IN CONTEXT: TECHNOLOGY, POLICY, AND THE INTEGRITY OF SOCIAL LIFE (2010); Daniel J. Solove, *A Taxonomy of Privacy*, 154 U. PA. L. REV. 477 (2006).
    86. Ryan Calo, *The Boundaries of Privacy Harm*, 86 IND. L.J. 1131, 1133 (2011).
    87. *Eightmaps.com and Too Much Information*, DALLAS MORNING NEWS (Jan. 14, 2009); *see also* Michael Shin, *Show Me the Money! The Geography of Contributions to California's Proposition 8*, 1 CAL. J. POL. & POL'Y 10 (2009). See generally on privacy-invasive online map services: Mark Burdon, *Privacy Invasive Geo-Mashups: Privacy 2.0 and the Limits of First Generation Information Privacy Laws*, 2010 U. ILL. J.L. TECH. & POL'Y 1 (2010).
    88. CAL. CONST. art. I, § 7.5 (enacted as California 2008 Ballot Proposition 8), *ruled unconstitutional in* Perry v. Schwarzenegger, 704 F.Supp.2d 921 (N.D. Cal. 2010).
    89. Deborah G. Johnson, Priscilla M. Regan & Kent Wayland, *Campaign Disclosure, Privacy and Transparency*, 19 WM. & MARY BILL RTS. J. 959, 972 (2011).
    90. *Id.*
    91. *Id. See also* Brad Stone, *Disclosure, Magnified on the Web*, N.Y. TIMES, Feb. 7, 2009, at BU3, http://www.nytimes.com/2009/02/08/business/08stream.html.

The feeling of having no control over one's personal information is a "subjective harm," described by Calo as "the perception of loss of control that results in fear or discomfort."[92] Many people are uncomfortable with organizations processing large amounts of information about them. Furthermore, there is often information asymmetry between the individual and the organization that uses personal information. People may know that information about them is collected and stored, but may not know how this will be used. If people do not know who holds data about them, they cannot exercise control over those data.[93] Releasing data to an undetermined number of re-users aggravates the lack of control.

Furthermore, data privacy rules that apply to the public sector are often stricter than those that apply to the private sector.[94] However, if the public sector releases personal data as open data, that is, with no restrictions, the private sector can subsequently use those data, subject to more lenient (statutory) rules.[95] Hence, releasing personal data as open data reduces privacy protection. Furthermore, the more datasets governments disclose, the richer the possibilities for re-identification. In sum, releasing personal information as open data causes a lack of individual control over personal information.

### 3. Social Sorting and Discrimination

A third privacy-related concern is that open data could be used as input for social sorting and discriminatory practices.[96] For instance, if the public sector released personal data, data brokers would likely be among the main re-users.[97] Data brokers are "companies that collect consumers' personal information and resell or share that information with others."[98] The information can be used, for instance, for direct marketing, credit scoring, or screening job applicants.[99]

---

92. Calo, *supra* note 86, at 1143.
93. *See generally* Alessandro Acquisti & Jens Grossklags, *What Can Behavioral Economics Teach Us About Privacy?*, in DIGITAL PRIVACY: THEORY, TECHNOLOGIES AND PRACTICES 363 (Sabrina De Capitani di Vimercati et al. eds., 2007); ZUIDERVEEN BORGESIUS, IMPROVING PRIVACY PROTECTION, *supra* note 73, at 201–05. According to Solove, the feeling of lost control resembles Franz Kafka's THE TRIAL. DANIEL J. SOLOVE, THE DIGITAL PERSON: TECHNOLOGY AND PRIVACY IN THE INFORMATION AGE 38 (2004). He suggests the main problem is "not knowing what is happening, having no say or ability to exercise meaningful control over the process." *Id.* at 38.
94. For instance, in the United States the Privacy Act of 1974 does not apply to the private sector. Pub. L. No. 93-579, 88 Stat. 1896 (codified at 5 U.S.C. § 552a (2012)). In the European Union, firms more easily meet the required legal basis test for personal data processing than public sector bodies do. *See* Directive 95/46/EC, *supra* note 69. Article 7(f) applies to firms; Article 7(e) applies to the public sector.
95. Scassa, *supra* note 71, at 405, 402.
96. *See, e.g.*, Solon Barocas & Andrew Selbst, *Big Data's Disparate Impact*, 104 CALIF. L. REV. (forthcoming 2016), http://ssrn.com/abstract=2477899; EXEC. OFFICE OF THE PRESIDENT, BIG DATA: SEIZING OPPORTUNITIES, PRESERVING VALUES 51 (2014), http://www.whitehouse.gov/sites/default/files/docs/big_data_privacy_report_may_1_2014.pdf; Scassa, *supra* note 71, at 407.

Many find data brokers' activities unfair and privacy-invasive.[100] As the Federal Trade Commission notes, personal information could be used for unfair discrimination. For instance, a company might use the information that there is a "Smoker in Household" to conclude that people in that household should not be offered insurance.[101] In surveillance studies, such practices are called "social sorting." As David Lyon explains, social sorting involves "obtain[ing] personal and group data in order to classify people and populations according to varying criteria, to determine who should be targeted for special treatment, suspicion, eligibility, inclusion, access, and so on."[102] Social sorting is not inherently bad or good.[103]

For social sorting, data brokers can also use open data that do not include personal information. For instance, the average housing price in a certain zip code is not personal information. But that average price could be matched with somebody's address to estimate the value of his or her house. Hence, non-personal information can be used to enrich digital dossiers about people.

The following is another example of a social sorting effect resulting from open data. Suppose a city council releases crime statistics. A vendor of GPS car systems can overlay its own maps with the crime data in order to designate high-crime areas. The car GPS system can then route the driver around those areas. The practice could be seen as unfair for the people and businesses in that newly invented no-go area. In the no-go areas, insurance premiums might rise, and real estate prices and shop profits might drop.

---

97. Thomas P. Keenan, *Are They Making Our Privates Public? Emerging Risks of Governmental Open Data Initiatives*, in PRIVACY AND IDENTITY MANAGEMENT FOR LIFE 1, 11 (Jan Camenisch et al. eds., 2012). *See also* Solove, *Access and Aggregation*, *supra* note 7, at 1148–50.

98. *See also* U.S. FEDERAL TRADE COMMISSION, DATA BROKERS: A CALL FOR TRANSPARENCY AND ACCOUNTABILITY 1 (2014), http://www.ftc.gov/system/files/documents/reports/data-brokers-call-transparency-accountability-report-federal-trade-commission-may-2014/140527databrokerreport.pdf.

99. *See* Scassa, *supra* note 71, at 407.

100. *See generally* Chris Jay Hoofnagle, *Big Brother's Little Helpers: How ChoicePoint and Other Commercial Data Brokers Collect and Package Your Data for Law Enforcement*, 29 N.C. J. INT'L L. & COM. REG. 595 (2003); JOSEPH TUROW, NICHE ENVY: MARKETING DISCRIMINATION IN THE DIGITAL AGE (2006); JOSEPH TUROW, THE DAILY YOU: HOW THE NEW ADVERTISING INDUSTRY IS DEFINING YOUR IDENTITY AND YOUR WORTH (2011).

101. FEDERAL TRADE COMMISSION, DATA BROKERS, *supra* note 98, at 55–56.

102. David Lyon, *Surveillance as Social Sorting: Computer Codes and Mobile Bodies*, in SURVEILLANCE AS SOCIAL SORTING: PRIVACY, RISK AND DIGITAL DISCRIMINATION 13 (David Lyon ed., 2002) [hereinafter Lyon, *Surveillance as Social Sorting*].

103. David Lyon, Kevin Haggerty & Kirstie Ball, *Introducing Surveillance Studies*, in ROUTLEDGE HANDBOOK OF SURVEILLANCE STUDIES 3 (Kirstie Ball, Kevin Haggerty & David Lyon eds., 2012).

In sum, potential privacy problems regarding open data include chilling effects on people communicating with the public sector, a lack of individual control over personal information, and discriminatory practices enabled by the released data. Hence, especially when datasets contain personal data, public sector bodies should give due consideration to the risks of disclosing data. We discuss below how to strike a balance between open data policy and privacy. But first we turn to the rules and guidelines that govern the disclosure of public sector information.

## III.   GOVERNANCE OF PUBLIC SECTOR INFORMATION

In this section we discusses governance frameworks regarding access to public sector information. We discuss norms that govern open data, and norms that govern access to public sector information more generally. Freedom of information laws provide inspiration on how to strike a balance between privacy and transparency in the open data context.

### A.   OPEN DATA NORMS

Obligations for public authorities to release information as open data tend not to be encoded in hard law. Rather, open data policy is often promoted through administrative hierarchies, whereby the policy objectives, targets, and instructions range from superficial and permissive to detailed and strict.[104] Open data policymaking is partly shaped through

---

104. For example, the 2013 order by President Obama breathes ambition and decisiveness, and the elaboration by the Office for Management & Budget of its Open Data Policy Memorandum contains specific duties for departments to create lists of available data sets ("Public Data Listing") and engage with user groups to prioritize release. OFFICE OF MGMT. & BUDGET, EXEC. OFFICE OF THE PRESIDENT, SUPPLEMENTAL GUIDANCE ON THE IMPLEMENTATION OF MEMORANDUM M-13-13 (2013), https://project-open-data.cio.gov/implementation-guide/; *see also* OMB MEMORANDUM M-13-13, OPEN DATA POLICY, *supra* note 25. The E.U.'s Public Sector Information Directive shows a preference for the release of data in open formats, and also demands that member states make practical arrangements "that help re-users in their search for documents available for re-use," e.g., in the form of asset registers. Directive 2003/98/EC, *supra* note 35. The E.C. Guidelines clearly favor pro-active release of data as open.

political commitment in international forums such as the G8 and the Open Government Partnership.[105]

Open data initiatives rely on norms that regulate access to information. After all, open government data are, by definition, publicly available data. A myriad of such norms exists at the national level. The most generic disclosure duties arise under freedom of information acts, which typically cover the executive branch. Constitutional and administrative norms that help cement basic checks and balances also have implications for access to information, mandating for example that legislative texts are published,[106] and that the public has access to court decisions.[107]

Additionally, many countries have dedicated laws that govern information production for specific purposes, such as (national) statistics to aid policy development and monitoring,[108] land registries to facilitate secure property transactions, business registers,[109] or earth observation data produced for environmental and agricultural management.[110] Such specific laws will often lay down modalities for access. For example, confidentiality of identifiable information is of fundamental interest for the production of reliable and useful statistics. Hence, a basic principle in instruments that govern the production and dissemination of statistics is that personal information supplied for statistical purposes will not be disclosed or used for other (administrative) purposes.¹¹¹ In the interest of research, some statistics offices organize secure environments, where researchers can access micro-data under strict conditions. While no international legal right to (re)use public sector information exists, access to government information is increasingly recognized as a human right.¹¹²

---

105. Through the *Open Data Charter*, the members of G8 have committed to drafting national open data action plans. *Supra* note 1. The same mechanism is used by the *Open Government Partnership*. *Supra* note 6.

106. *See, e.g.*, 1958 CONST. arts. 10–11 (Fr.); GRUNDGESETZ FÜR DIE BUNDESREPUBLIK DEUTSCHLAND [GRUNDGESETZ] [GG] [BASIC LAW], May 23, 1949, BGBL. I, Art. 82 (Ger.).

107. For example, Article 6 of the European Convention on Human Rights (on the right to a fair trial) prescribes that court decisions are to be made public. ECHR, *supra* note 67, art. 6. This will usually be through delivery in court but may be achieved by other means as well. *See* COUNCIL OF EUROPE, GUIDE ON ARTICLE 6, at 49–50 (2013). That a right to information is no guarantee for easy and affordable access is witnessed by the electronic access system for federal courts. *See* Vera Eidelman & Amul Kalia, *Right to Know: The PACER Mess and How to Clean It*, ELEC. FRONTIER FOUND. (Sept. 2, 2014). https://www.eff.org/deeplinks/2014/09/right-know-pacer-mess-and-how-clean-it.

108. *See, e.g.*, Stb. 2003, p. 551 [Act on the Central Bureau of Statistics] (Neth.); Statistics Act, R.S.C. 1985, c. S-19 (Can.); Statistics and Registration Service Act, 2007, c. 18 (U.K.).

109. *See, e.g.*, Stb. 2007, p. 153 [Act on Trade Register] (Neth.); Companies Act, 2006, c. 46 (U.K.); Handelsgesetzbuch [Act on Trade Register], Oct. 23, 2008, BGBL. III at 4100-1, § 8 (Ger.).

110. Mireille van Eechoud, *Commercialization of Public Sector Information: Delineating the Issues*, *in* THE FUTURE OF THE PUBLIC DOMAIN: IDENTIFYING THE COMMONS IN INFORMATION LAW 279, 281–83 (Lucie Guibault & P. Bernt Hugenholtz eds., 2006).

B.   ACCESS TO INFORMATION NORMS

Several international courts see access rights as part of, or closely connected to, the right to freedom of expression.¹¹³ However, access rights are also recognized in case law of the European Court of Human Rights in the context of the right to private life.[114] By contrast, access rights may be conceived of as stand-alone constitutional rights.[115]

---

111. *See* CONFERENCE OF EUROPEAN STATISTICIANS, FUNDAMENTAL PRINCIPLES OF OFFICIAL STATISTICS (1991). They were since updated and endorsed by the U.N. General Assembly. G.A. Res. 68/261, U.N. Doc. A/68/261 (Jan. 29, 2014). Principle 6 reads: "Individual data collected by statistical agencies for statistical compilation, whether they refer to natural or legal persons, are to be strictly confidential and used exclusively for statistical purposes." *Id.* There are similar examples at the national level. *See, e.g.*, Confidential Information Protection and Statistical Efficiency Act of 2002 (CIPSEA), Pub. L. 107-347, 116 Stat. 2962 (2002); Stb. 2003, p. 516 [Dutch Statistics Act] (Neth.).

112. For an extensive analysis of different human rights based conceptualizations of access to government information, see CHERYL A. BISHOP, ACCESS TO INFORMATION AS A HUMAN RIGHT (2011).

113. A right to access of government information is guaranteed under Article 13 (Freedom of Thought and Expression) of the American Convention on Human Rights, and states have a positive obligation to provide access, subject only to access restrictions that are proportionate and for reasons permitted by the Convention. Claude-Reyes et al. v. Chile, Merits, Reparations, and Costs, Judgment, Inter-Am. Ct. H.R. (ser. C) No. 151, ¶ 77 (Sept. 9, 2006). Refusal to grant access to government information to a public watchdog violates the right to freedom of expression under Article 10 of the ECHR. Youth Initiative for Human Rights v. Serbia, App. No. 48135/06, Judgment, 2013 Eur. Ct. H.R. 584 (2013); ECHR, *supra* note 67, art. 10. In another case, the European Court of Human Rights (ECtHR) conceded it "has recently advanced towards a broader interpretation of the notion of 'freedom to receive information' and thereby towards the recognition of a right of access to information." TASZ v. Hungary, App. No. 37374/05, 2009 Eur. Ct. H.R., ¶ 35 (2009). Previously it had rejected the claim that ECHR Article 10 includes a right to access government information, or a positive obligation for states to collect and disseminate information. *See, e.g.*, Guerra v. Italy, App. No. 14967/89, 1998 Eur. Ct. H.R. 7 (1998).

The Tromsø Convention of the Council of Europe concerns access to government information,[116] but it is unlikely that enough member states will ratify this convention for it to enter into force any time soon.[117] Much more successful is the U.N. Aarhus Convention of 1998, with nearly fifty contracting states.[118] The Aarhus Convention provides for a right of access to environmental information as part of every citizen's right to an adequate environment and duty to safeguard the environment for future generations.[119]

A fundamental right of access does not necessarily imply that authorities must actively disclose information to the general public in electronic form without use-restrictions. But open government agendas do steer policy in that direction. At the global level, the Open Government Partnership promotes proactive disclosure in reusable formats.[120]

In various human rights domains, proactive disclosure is also advocated. The U.N. rapporteur on Human Rights typifies the right to access government information as "one of the central components of the right to freedom of opinion and expression."[121] To give effect to the right

---

114. The ECtHR recognized a duty to impart information for the government as part of the right to respect for private life (under Article 8 of the ECHR) on various occasions: where it concerned access to foster care records, Gaskin v. UK, App. No. 10454/83, 12 Eur. H.R. Rep. 36 (1989), and with respect to information about environmental pollution (threatening citizens' health), Guerra v. Italy, *supra* note 113; Onderyildiz v. Turkey, 2004-XII Eur. Ct. H.R. 81. In these cases, applicants had a special interest.

115. For example, Article 42 of the Charter of Fundamental Rights of the European Union provides that any citizen of the Union has a right of access to documents held by E.U. institutions. E.U. Charter of Fundamental Rights, *supra* note 68, art. 42. For an in depth analysis of access rights of a wider openness agenda, see Alberto Alemanno, *Unpacking the Principle of Openness in EU Law: Transparency, Participation and Democracy*, 39 EUR. L. REV. 72 (2014).

116. Tromsø Convention, Council of Europe Convention on Access to Official Documents, *opened for signature* June 18, 2009, C.E.T.S. No. 205 (not yet ratified).

117. Mireille van Eechoud & Katleen Janssen, *Rights of Access to Public Sector Information*, 6 MASARYK U. J.L. & TECH. 471, 486 (2012).

118. Aarhus Convention on Access to Information, Public Participation in Decision-Making and Access to Justice in Environmental Matters, *opened for signature* June 25, 1998, 2161 U.N.T.S. 447 (entered into force Oct. 30, 2001).

119. *Id.*

120. *See Open Government Declaration*, OPEN GOVERNMENT PARTNERSHIP, http://www.opengovpartnership.org/about/open-government-declaration (last visited May 1, 2015).

121. Special Rapporteur, *Report on the Promotion and Protection of the Right to Freedom of Opinion and Expression in Accordance with Human Rights Council Resolution 16/4,* of access to information under Article Nineteen of the United Nations International Covenant on Civil and Political Rights and the Universal Declaration of Human Rights, "parties should proactively put in the public domain Government information of public interest" and "make every effort to ensure easy, prompt, effective and practical access to such information."[122] In 2006, the Inter-American Court of Human Rights held that States have a positive obligation to legislate freedom of information laws or take other measures that ensure access to government information.[123]

The adoption rate of freedom of information laws has accelerated on all continents over the past decade. Today nearly a hundred countries have enacted freedom of information laws.[124] Some freedom of information laws contain provisions on proactive disclosure of information.[125] These tend to be vague and rather limited in scope. Traditionally access laws focus on disclosure of information on request by a member of the public. Access laws detail how requests can be made and how decisions must be reached.[126] A basic principle in freedom of information acts is that citizens do not have to motivate why they want access; the public interest in

---

*transmitted by Note of the Secretary-General*, U.N. Doc. A/68/362 (Sept. 4, 2013). *See also* Human Rights Council Res. 12/12, Right to the Truth, 12th Sess., Oct. 1, 2009, U.N. GAOR, 68th Sess., A/HRC/RES/12/12, at 3 (Oct. 12, 2009) ("the public and individuals are entitled to have access, to the fullest extent practicable, to information regarding the actions and decision-making processes of their Government, within the framework of each State's domestic legal system"); Inter-American Commission on Human Rights [IACHR], *The Right to Truth in the Americas*, IACHR Doc. OEA/Ser.L/V/II.152 (Aug. 13, 2014).

122. Human Rights Comm. on Article 19: Freedoms of Opinion and Expression, General Comment No. 34, Rep. on its 102d Sess., July 11–29, 2011, U.N. Doc. CCPR/C/GC/34, ¶ 19 (Sept. 12, 2011).

123. Claude-Reyes et al. v. Chile, Merits, Reparations, and Costs, Judgment, Inter-Am. Ct. H.R. (ser. C) No. 151, ¶¶ 77, 102 (Sept. 9, 2006).

124. *See* Map, GLOBAL RIGHT TO INFORMATION RATING, http://www.rti-rating.org.

125. For an analysis of the drivers of pro-active disclosure of government information and its growing enactment in binding norms, see Helen Darbishire, *Proactive Transparency: The Future of the Right to Information?* (World Bank Inst. Governance Working Paper Series, No. 56,598, 2010).

126. *See* JONATHAN GRAY & HELEN DARBISHIRE, BEYOND ACCESS: OPEN GOVERNMENT DATA & THE RIGHT TO (RE)USE PUBLIC INFORMATION (Creative Commons, 2011); Mireille van Eechoud et al., *Good Practices Collection on Access to Data*, LAPSI (July 11, 2014) [hereinafter *Good Practices Collection*].

disclosure is considered a given.[127] A right to access information does not necessarily imply that the information can subsequently be used freely.[128]

Generally, freedom of information laws do not prescribe how data must be made available (for example in an open format, machine readable, with a certain frequency).[129] Usually, information disclosed under freedom of information laws is not required to be legally or technically open.[130] It is, however, a common feature that public bodies must, wherever possible, respect the mode of supply preferred by the requesting party, if the documents are available in such form or easily so produced.[131] Freedom of information laws usually contain privacy provisions, as discussed next.

C.    ACCESS TO INFORMATION NORMS AND PRIVACY

Machine readable, bulk-downloadable open data complicate a problem that was already a difficult one in the pre-digital era. Since at least the 1970s, countries have grappled with the problem of balancing privacy protection and public sector transparency.[132] Generic freedom of information laws typically aim to accommodate privacy interests, for example by reserving access to personal information to parties with particular interests, or by only making records available in secure reading rooms.

Two balancing models regarding privacy and transparency can be distinguished in freedom of information laws. First, sometimes privacy is an absolute limitation to disclosure. That is, the legislator has done the balancing ex-ante. For example, the Dutch Freedom of Information Act provides that certain types of sensitive personal data (for example data concerning medical matters or religion) may never be disclosed.[133]

---

127. GRAY & DARBISHIRE, *supra* note 126; *Good Practices Collection*, *supra* note 126.
128. For instance, before implementation of the E.U. Public Sector Information Directive, the Belgian federal freedom of information act stipulated that no commercial use was allowed of information obtained under the act. *See* Wet betreffende de openbaarheid van bestuur of Apr. 11, 1994, BELGISCH STAATSBLAD [B.S.] [Official Gazette of Belgium], June 30, 1994 (deleted by Act N. 2007-1600, Mar. 7, 2007).
129. See the analysis of over forty freedom of information acts, GRAY & DARBISHIRE, *supra* note 126; *Good Practices Collection*, *supra* note 126.
130. GRAY & DARBISHIRE, *supra* note 126.
131. *See, e.g.*, Aarhus Convention of 1998, *supra* note 118, art. 4.
132. For instance, in 1973 Sweden adopted its data privacy law partly to ensure that the generous Swedish regime for access to official documents, which dates back to 1776, would not unduly interfere with privacy. *See* GLORIA GONZÁLEZ FUSTER, THE EMERGENCE OF PERSONAL DATA PROTECTION AS A FUNDAMENTAL RIGHT OF THE EU 59 (2014).
133. Wet openbaarheid van bestuur [Dutch Freedom of Information Act], Stb. 1991, art. 10(1)d (Neth.).

Second, sometimes freedom of information laws include a relative privacy exemption, to be weighed against the public interest in disclosure on a case-by-case basis.[134] U.S. freedom of information law exempts disclosure of personal, medical and similar files.[135] The test is whether disclosure "would constitute a clearly unwarranted invasion of personal privacy."[136] Personal information gathered as part of law enforcement is also exempt, if disclosure "could reasonably be expected to constitute an unwarranted invasion of personal privacy."[137] If privacy interests prevent disclosure, it is common for freedom of information laws to demand that exempted information is redacted so that the remainder can be released, even if cleaning documents is labor intensive.[138]

The regulation that governs access to documents from E.U. institutions (such as the Council of Ministers, Parliament, and Commission) stipulates that access to a document shall be refused if "disclosure would undermine the protection of privacy . . . in particular in accordance with Community legislation regarding the protection of personal data."[139] The Obama Freedom of Information Memorandum states: "In the face of doubt, openness prevails."[140]

At global human rights forums, the presumption is that the public interest in access to public sector information (as part of the freedom of expression) trumps privacy and other interests. Human rights rapporteurs for the United Nations argue that access to information should be granted unless disclosure would cause serious harm to a protected interest such as privacy that outweighs the interest in disclosure.[141] The rapporteurs also stress the importance of proactive disclosure obligations, and posit that "access to information law should, to the extent of any inconsistency, prevail over other legislation."[142]

---

134. The Dutch Freedom of Information Act provides such a relative ground for non-disclosure, where the public's right to know does not outweigh a person's interest to have his or her private sphere protected. *Id.*, art. 10(2)e.

135. 5 U.S.C. § 552(b)(6) (2012) (known as Exemption 6 of Electronic Freedom of Information Act of 1966, Pub. L. No. 104-231, 110 Stat. 3048).

136. *Id.*

137. 5 U.S.C. § 552(b)(7).

138. For example, Article 4(4) of the Aarhus Convention exempts the release of personal data (if confidential under domestic law); Article 4(6) obliges states to redact the documents. Aarhus Convention of 1998, *supra* note 118, art. 4.

139. Council Regulation 1049/2001, art. 4, 2001 O.J. (L 145) 43. The way the institutions have interpreted this limitation is controversial; the European Ombudsman and the European Data Protection Supervisor signal overzealous interpretation of the rules on data protection as a threat to transparency. How the scales tip thus depends as much on the prevailing culture of transparency (or secrecy) as on the black letter. *See* H. R. KRANENBORG, TOEGANG TOT DOCUMENTEN EN BESCHERMING VAN PERSOONSGEGEVENS IN DE EUROPESE UNIE [ACCESS TO DOCUMENTS AND DATA PROTECTION IN THE EUROPEAN UNION] 188–94 (2007).

140. *See* President Barack Obama, Memorandum, Freedom of Information Act, 74 Fed. Reg. 15 (Jan. 21, 2009).

Particularly for the disclosure in the interest of political accountability and public debate, judgments in which the right to freedom of expression and the right to privacy are balanced can give guidance. The European Court of Human Rights recognizes the importance of proactive release of data on the Internet as a means to ensure effective transparency and accountability. In the *Wypych* case, the Court rejected the claim by an elected local councilor who argued that by requiring him to disclose information on his financial interests online, the Polish legislature infringed his right to privacy under Article 8 of the European Convention on Human Rights. The Court noted "[t]he general public has a legitimate interest in ascertaining that local politics are transparent and Internet access to the declarations makes access to such information effective and easy. Without such access, the obligation would have no practical importance or genuine incidence on the degree to which the public is informed about the political process."[143]

Earlier, the European Court of Human Rights held that the privacy interests of politicians and higher public officials must yield to access rights.[144] The Court considered "that it would be fatal for freedom of expression in the sphere of politics if public figures could censor the press and public debate in the name of their personality rights, alleging that their opinions on public matters are related to their person and therefore constitute private data which cannot be disclosed without consent."[145]

---

141. *See, e.g.*, AMBEYI LIGABO, U.N. SPECIAL RAPPORTEUS ON FREEDOM OF OPINION AND EXPRESSION ET AL., JOINT DECLARATION, INTERNATIONAL MECHANISMS FOR PROMOTING FREEDOM OF EXPRESSION (2006) [hereinafter JOINT DECLARATION, PROMOTING FREEDOM OF EXPRESSION (2006)] (signed by the U.N. Special Rapporteur on Freedom of Opinion and Expression, the OSCE Representative on Freedom of the Media, the OAS Special Rapporteur on Freedom of Expression and the African Commission on Human and Peoples' Rights). *See also* JOINT DECLARATION, INTERNATIONAL MECHANISMS FOR PROMOTING FREEDOM OF EXPRESSION (2004) [hereinafter JOINT DECLARATION, PROMOTING FREEDOM OF EXPRESSION (2004)].
142. *See, e.g.*, JOINT DECLARATION, PROMOTING FREEDOM OF EXPRESSION (2004), *supra* note 141.
143. Wypych v. Poland, App. No. 2428/05, 2005 Eur. Ct. H.R. (admissibility decision).
144. TASZ v. Hungary, *supra* note 113, ¶ 37.
145. TASZ v. Hungary, *supra* note 113, ¶ 37.

In conclusion, regulation and case law regarding freedom of public sector information can provide inspiration on how to strike the balance between privacy and transparency in the open data context. Apart from that, there are more general principles to balance privacy-related interests and other interests. We turn to those Fair Information Principles now.

IV. GOVERNANCE OF PERSONAL INFORMATION

The Fair Information Principles (FIPs) provide a framework to balance privacy and other interests. Below we give an introduction to the FIPs, and to the OECD Privacy Guidelines, which include a version of the FIPs. We also discuss the main challenges when reconciling the FIPs and open data policy.

A. FAIR INFORMATION PRINCIPLES (FIPS)

1. *Background of the FIPs*

The Fair Information Principles (FIPs),[146] or the Fair Information Practice Principles (FIPPs),[147] are ingrained in most data privacy laws and guidelines around the world. For example, the FIPs can be recognized in the 1973 report *Records, Computers, and the Rights of Citizens*, by the U.S. Department of Health, Education, and Welfare;[148] the Privacy Act;[149] and the Fair Credit Reporting Act.[150] The Federal Trade Commission and the White House have recently called for FIPs-based privacy regulation for the private sector.[151]

---

146. *See* NEIL RICHARDS, INTELLECTUAL PRIVACY: RETHINKING CIVIL LIBERTIES IN THE DIGITAL AGE 162 (2014) *See also* Robert Gellman, *Fair Information Practices: A Basic History, Version 2.02*, BOBGELLMAN.COM (2013), http://bobgellman.com/rg-docs/rg-FIPShistory.pdf.

147. *See The Fair Information Principles at Work*, U.S. DEP'T OF HOMELAND SECURITY, http://www.dhs.gov/xlibrary/assets/privacy/dhsprivacy_fippsfactsheet.pdf.

148. U.S. DEP'T OF HEALTH, EDUC. & WELFARE, RECORDS, COMPUTERS, AND THE RIGHTS OF CITIZENS, at i, xx (1973), http://www.justice.gov/opcl/docs/rec-com-rights.pdf.

149. Privacy Act of 1974, Pub. L. No. 93-579, 88 Stat. 1896 (codified at 5 U.S.C. § 552a (2012)).

150. Fair Credit Reporting Act of 1970, Pub. L. No. 91-508, 84 Stat. 1128 (codified as amended at 15 U.S.C. §§ 1681–1681x (2012)).

151. OFFICE OF THE PRESIDENT, CONSUMER DATA PRIVACY IN A NETWORKED WORLD (2012), http://www.whitehouse.gov/sites/default/files/privacy-final.pdf; FEDERAL TRADE COMMISSION, PROTECTING CONSUMER PRIVACY IN AN ERA OF RAPID CHANGE: RECOMMENDATIONS FOR BUSINESSES AND POLICYMAKERS (2012), http://www.ftc.gov/sites/default/files/documents/reports/federal-trade-commission-report-protecting-consumer-privacy-era-rapid-change-recommendations/120326privacyreport.pdf.

About a hundred countries in the world have a data privacy law including a version of the FIPs.[152] The FIPs can also be recognized in the United Nations Guidelines for the Regulation of Computerized Personal Data Files 1990,[153] and the APEC Privacy Framework of the Asia-Pacific Economic Cooperation (2005).[154] The E.U. Data Protection Directive (1995) contains one of the world's most stringent implementations of the FIPs.[155] European legal scholars tend to speak of data protection principles rather than of FIPs, but both sets of principles are similar.[156] Different countries, however, implement the FIPs differently. The FIPs give guidelines to balance privacy-related interests and other interests, such as those of business and the public sector.[157]

### 2. OECD Guidelines

An influential version of the FIPs can be found in the *Guidelines Governing the Protection of Privacy and Transborder Flows of Personal Data*, from the Organisation for Economic Co-operation and Development (OECD).[158] The OECD was established in 1960, by eighteen European countries, the United States, and Canada.[159] Now, the OECD has thirty-four member countries, including Mexico, Chile, Korea and Japan.[160] The OECD's self-stated mission is "to promote policies that will improve the economic and social well-being of people around the world."[161]

---

152. Graham Greenleaf, *Sheherezade and the 101 Data Privacy Laws: Origins, Significance and Global Trajectories*, 23 J.L. INFO. & SCI. 4 (2014); GRAHAM GREENLEAF, GLOBAL TABLES OF DATA PRIVACY LAWS AND BILLS (3d ed. 2013), http://ssrn.com/abstract=2280875.

153. Guidelines for the Regulation of Computerized Personal Data Files, G.A. RES. 45/95, U.N. DOC. A/RES/45/95 (Dec. 14, 1990).

154. Asia-Pacific Economic Cooperation [APEC], *Privacy Framework*, APEC Doc. No. 205-SO-01.2 (2005), http://www.apec.org/Groups/Committee-on-Trade-and-Investment/~/media/Files/Groups/ECSG/05_ecsg_privacyframewk.ashx.

155. Directive 95/46/EC, *supra* note 69.

156. The core of E.U. data protection law can be found in article 6 of the Data Protection Directive. Directive 95/46/EC, *supra* note 69, art. 6.

157. Paul de Hert & Serge Gutwirth, *Privacy, Data Protection and Law Enforcement: Opacity of the Individual and Transparency of Power*, in PRIVACY AND THE CRIMINAL LAW 91 (Erik Claes, Antony Duff, & Serge Gutwirth eds., 2006); *see also* RICHARDS, *supra* note 146, at 162; Marc Rotenberg, *Fair Information Practices and the Architecture of Privacy (What Larry Doesn't Get)*, 2001 STAN. TECH. L. REV. 1, 1–4; Ann Cavoukian, *Evolving FIPPs: Proactive Approaches to Privacy, Not Privacy Paternalism*, in REFORMING EUROPEAN DATA PROTECTION LAW 293 (Serge Gutwirth, Ronald Leenes & Paul de Hert eds., 2015).

158. *OECD Guidelines on the Protection of Privacy and Transborder Flows of Personal Data*, OECD, http://www.oecd.org/sti/ieconomy/oecdguidelinesontheprotectionofprivacyandtransborderflowsofpersonaldata.htm (last visited June 22, 2015) [hereinafter OECD Privacy Guidelines]. The OECD Privacy Guidelines call these principles "Basic Principles of National Application."

159. Robert Wolfe, *From Reconstructing Europe to Constructing Globalization: The OECD in Historical Perspective*, in THE OECD AND TRANSNATIONAL GOVERNANCE 25, 25–26 (Rianne Mahon & Stephen McBride eds., 2008).

One of the main reasons for the OECD to adopt the Guidelines was that several European data privacy laws from the 1970s restricted the export of personal data to countries that offered inadequate legal protection to personal data. Some, the United States in particular, worried that European countries would use data privacy law as a trade barrier.[162] The chairman of the expert group that wrote the 1980 OECD Guidelines summarized, "the OECD's central concern was . . . that the response of European nations (and European regional institutions) to the challenges of TBDF [transborder data flows] for privacy might potentially erect legal and economic barriers against which it was essential to provide effective exceptions."[163] Therefore, OECD member states negotiated about more international cooperation, leading to the adoption of the Privacy Guidelines in 1980.[164]

The OECD Guidelines have a dual goal: they aim to protect privacy and individual liberties, and to foster the free flow of information between OECD member countries.[165] Many legal data privacy instruments have a similar dual goal.[166] In this Article we focus on protecting privacy and individual liberties, rather than on transborder data flows.[167]

The Guidelines are not legally binding, they merely "recommend" that OECD member countries implement the Guidelines.[168] The Guidelines stress that they provide "minimum standards"[169] and do not "preven[t] the application of different protective measures to different categories of personal data, depending upon their nature and the context in which they are collected, stored, processed or disseminated."[170] The OECD Guidelines use flexible terms so that all of the member countries can agree with them, even though the United States and European countries have different legal traditions, especially regarding privacy and personal data.[171]

---

160. *Members and Partners*, OECD, http://www.oecd.org/about/membersand partners (last visited June 22, 2015).

161. *About the OCED*, OECD, http://www.oecd.org/about (last visited June 22, 2015).

162. Nicholas Platten, *Background to and History of the Directive*, in EC DATA PROTECTION DIRECTIVE 15 (David Bainbridge ed., 1996); GONZÁLEZ FUSTER, *supra* note 132, at 77.

163. Michael Kirby, *The History, Achievement and Future of the 1980 OECD Guidelines on Privacy*, 20 J.L. INFO. & SCI. 1, 6 (2010).

164. *Id.* at 7–10.

165. OECD Privacy Guidelines, *supra* note 158.

166. *See* GONZÁLEZ FUSTER, *supra* note 132, at 130. For instance, the E.U. Data Protection Directive has a similar dual goal. *See* Directive 95/46/EC, *supra* note 69, art. 1.

167. On transborder data flows, see CHRISTOPHER KUNER, TRANSBORDER DATA FLOWS AND DATA PRIVACY LAW (2013).

168. OECD, *Recommendation of the Council Concerning Guidelines Governing the Protection of Privacy and Transborder Flows of Personal Data*, at 11–12, C(80)58/FINAL (2013), http://www.oecd.org/sti/ieconomy/2013-oecd-privacy-guidelines.pdf (as amended on July 11, 2013).

When the OECD Guidelines were adopted in 1980, only about one third of the member states had adopted a data privacy law. Now, almost every OECD member state has a data privacy law with the FIPs at its core.[172] The OECD Guidelines were updated in 2013, but the essence of the principles was retained.[173] The 2013 OECD Privacy Guidelines are listed below. The principles partly overlap, and should be read together:

> **Collection Limitation Principle**
> There should be limits to the collection of personal data and any such data should be obtained by lawful and fair means and, where appropriate, with the knowledge or consent of the data subject.[174]
>
> **Data Quality Principle**
> Personal data should be relevant to the purposes for which they are to be used, and, to the extent necessary for those purposes, should be accurate, complete and kept up-to-date.[175]
>
> **Purpose Specification Principle**
> The purposes for which personal data are collected should be specified not later than at the time of data collection and the subsequent use limited to the fulfilment of those purposes or such others as are not incompatible with those purposes and as are specified on each occasion of change of purpose.[176]

---

169. *Id.* at 14.
170. *Id.* at 13.
171. Kirby, *supra* note 163, at 10.
172. David Wright, Paul de Hert, & Serge Gutwirth, *Are the OECD Guidelines at 30 Showing Their Age?*, 54 COMMUNICATIONS OF THE ACM 119, 122 (2011).
173. OECD, THE OECD PRIVACY FRAMEWORK 4, http://www.oecd.org/sti/ieconomy/oecd_privacy_framework.pdf (last visited June 22, 2015) ("[T]his revision leaves intact the original 'Basic Principles' in Part Two of the Guidelines.").
174. *Id.* at 14 (paragraph 7 of the Guidelines governs the protection of privacy and transborder flows of personal data).
175. *Id.*
176. *Id.*

**Use Limitation Principle**
> Personal data should not be disclosed, made available or otherwise used for purposes other than those specified in accordance with [the Purpose Specification Principle] except:
> > a) with the consent of the data subject; or
> > b) by the authority of law.[177]

**Security Safeguards Principle**
> Personal data should be protected by reasonable security safeguards against such risks as loss or unauthorised access, destruction, use, modification or disclosure of data.[178]

**Openness Principle**
> There should be a general policy of openness about developments, practices and policies with respect to personal data. Means should be readily available of establishing the existence and nature of personal data, and the main purposes of their use, as well as the identity and usual residence of the data controller.[179]

**Individual Participation Principle**
> Individuals should have the right:
> > a) to obtain from a data controller, or otherwise, confirmation of whether or not the data controller has data relating to them;
> > b) to have communicated to them, data relating to them
> > > (i) within a reasonable time;
> > > (ii) at a charge, if any, that is not excessive;
> > > (iii) in a reasonable manner; and
> > > (iv) in a form that is readily intelligible to them;
> > c) to be given reasons if a request made under subparagraphs (a) and (b) is denied, and to be able to challenge such denial; and
> > d) to challenge data relating to them and, if the challenge is successful to have the data erased, rectified, completed or amended.[180]

---

177. *Id.*
178. *Id.* at 15.
179. *Id.*
180. *Id.*

**Accountability Principle**
> A data controller should be accountable for complying with measures which give effect to the principles stated above.[181]

### 3. Scope of the OECD Guidelines

The OECD Guidelines apply to "personal data," which the Guidelines define as "any information relating to an identified or identifiable individual (data subject)."[182] But the Guidelines limit the scope of application considerably; they "apply to personal data, whether in the public or private sectors, which, *because of the manner in which they are processed, or because of their nature or the context in which they are used, pose a risk to privacy and individual liberties.*"[183]

The Guidelines thus follow a risk-based approach: they only apply to personal data processing if it threatens privacy or individual liberties. By contrast, E.U. data protection law generally applies to personal data processing, and requires that personal data be processed fairly, including when the data do not pose a prima facie risk for individual liberties.[184]

In this Article, we assume that personal data should always be handled in line with the FIPs.[185] Hence, we do not follow the risk-based approach suggested by the OECD Guidelines. We do consider the risk of personal data processing and the sensitivity of personal data, but we do so *within* the FIPs framework (*see infra* Parts V–VII).

The OECD Guidelines have been criticized, for instance, for implementing the FIPs too weakly. Roger Clarke says the OECD Guidelines aim "to facilitate international business, *not* to protect privacy."[186] The OECD Guidelines "were motivated by the facilitation of

---

181. *Id.*
182. *Id.* at 13. The OECD personal data definition is similar to the definition in E.U. data protection law. Directive 95/46/EC, *supra* note 69, art. 2(a).
183. OECD PRIVACY FRAMEWORK, *supra* note 173, at 14 (emphasis added).
184. *See* E.U. Charter of Fundamental Rights, *supra* note 68, art. 8 ("1. Everyone has the right to the protection of personal data concerning him or her. 2. Such data must be processed fairly for specified purposes and on the basis of the consent of the person concerned or some other legitimate basis laid down by law."); *see also* Case C-131/12, Google Inc. v Agencia Española de Protección de Datos (AEPD), 2014 EUR-Lex CELEX LEXIS 0131 ¶ 69 (May 13, 2014) (CJEU); Joined Cases 293 & 594/12, Digital Rights Ireland Ltd v Minister for Communications, Marine and Natural Resources, Kärntner Landesregierung, 2014 EUR-Lex CELEX LEXIS 0293 ¶ 36 (Apr. 8, 2014) (CJEU).
185. The idea that personal data should always be processed in line with the FIPs could be seen as a European approach.
186. Roger Clarke, Research Use of Personal Data, Comments at the National Scholarly Communications Forum on Privacy: Balancing the Needs of Researchers and international business; they were constrained by the need to leave existing legislation unaffected; and their formulation reflected the need for cross-cultural comprehensibility."[187]

For better or for worse, the FIPs are widely accepted as a starting point for data privacy law. Although the application of the FIPs varies considerably, they express a nearly worldwide consensus on minimum standards for fair personal data use. The next section describes the main challenges that arise when trying to reconcile the FIPs and open data policy.

B.     FIPs AND OPEN DATA: CHALLENGES

To date, policymakers and academics have given limited attention to the question of how privacy norms might be reconciled with policies aimed at making government data available for a wide range of uses. Policymakers and civil society actors recognize the privacy implications of open data.[188] But detailed analyses of the tension between open data and privacy, and especially of open data and the FIPs, is scarce. In the empirical mapping study, we found that, while there were some mentions of open data and privacy together on various forms of digital media, many of these were fleeting or incidental, and few of them contained substantive discussion about how to achieve a balance between the two.[189]

Several open data guidelines from civil society mention privacy—albeit cursorily. For example, the *8 Principles of Open Government Data* state that "[r]easonable privacy, security and privilege restrictions may be allowed."[190] The Sunlight Foundation says that for a dataset to be open, "[a]ll raw information from [the] dataset should be released to the public, except to

---

the Individual's Right to Privacy under the New Privacy Laws (Aug. 9, 2002), http://www.rogerclarke.com/DV/NSCF02.html; *see also* William Bonner & Mike Chiasson, *If Fair Information Principles Are the Answer, What Was the Question? An Actor–Network Theory Investigation of the Modern Constitution of Privacy*, 15 INFO. & ORG. 267, 284 (2005).

187. Roger Clarke, *Beyond the OECD Guidelines: Privacy Protection for the 21st Century* (Jan. 4, 2000), http://www.rogerclarke.com/DV/PP21C.html.

188. For example, the United Kingdom's Open Rights Group expressed concern over the U.K. government's plans to release anonymized health and education data. *See Open Data Privacy*, OPEN RIGHTS GROUP, https://www.openrightsgroup.org/campaigns/opendata/open-data-privacy (last visited June 22, 2015). The Open Knowledge and the Open Rights Group convened a working group on open data, personal data, and privacy. *See* PERSONAL DATA & PRIVACY WORKING GROUP, http://personal-data.okfn.org/ (last visited June 22, 2015).

189. MAPPING THE POLITICS OF OPEN DATA, *supra* note 62.

190. *8 Principles of Open Government Data*, PUBLIC.RESOURCE.ORG (Dec. 8, 2007), https://public.resource.org/8_principles.html.

the extent necessary to comply with federal law regarding the release of personally identifiable information."[191]

In the open data context, governmental and intergovernmental bodies also mention protecting privacy, albeit in a cursory fashion. For example, the *G8 Open Data Charter* recognizes that "there is national and international legislation, in particular pertaining to intellectual property, personally-identifiable and sensitive information, which must be observed."[192] The 2008 OECD recommendation on public sector information urges that member countries should clearly define "grounds of refusal or limitations," including "personal privacy."[193]

Compared to the OECD's recommendation, the implementation guidance material for Obama's 2013 executive order contains a more substantive discussion of privacy and the Fair Information Principles. The memorandum includes the suggestion to "[s]trengthen measures to ensure that privacy and confidentiality are fully protected and that data are properly secured[,]" and to "incorporate privacy analyses into each stage of the information's life cycle."[194] As well as demanding compliance with relevant laws such as the U.S. Privacy Act of 1974 and the E-Government Act of 2002, the memorandum suggests that "agencies should implement information policies based upon Fair Information Practice Principles and NIST guidance on Security and Privacy Controls for Federal Information Systems and Organizations."[195]

In the European Union, some work has been done on reconciling privacy and open data, in a thematic network funded by the European Commission to reflect on Legal Aspects of Public Sector Information (LAPSI). The LAPSI Working Group on privacy warns that full application of European data privacy rules will seriously hamper the ability of public sector bodies to disclose information for re-use purposes.[196] In the following section, we discuss the main challenges that occur when trying to reconcile the FIPs and open data policy, starting with the purpose specification principle.

---

191. SUNLIGHT FOUNDATION, *supra* note 12.
192. G8 OPEN DATA CHARTER, *supra* note 1.
193. *OECD Recommendation*, *supra* note 19, at 5.
194. OMB MEMORANDUM M-13-13, OPEN DATA POLICY, *supra* note 25, at 9.
195. *Id.*
196. Dos Santos et al., *supra* note 8, at 348–49; *see also* van Eechoud et al., *LAPSI Position Paper*, *supra* note 8.

### 1. *Purpose Specification Principle*

The main problem that occurs when trying to reconcile the FIPs and open data policy is that open data policy fosters unanticipated re-use and innovation—"serendipitous reuse" as Shadbolt et al. put it.[197] But secondary use of personal data brings privacy risks. In FIPs parlance, using personal information for unforeseen purposes may breach the purpose specification principle.

The purpose specification principle is a cornerstone of many data privacy laws in the world. It follows from the purpose principle that personal data should only be collected for a purpose that is specified in advance, and that those data should not be used for incompatible purposes.[198] The 1973 "Records, Computers, and the Rights of Citizens" report from the U.S. Department of Health, Education, and Welfare already contained a similar principle: "[t]here must be a way for an individual to prevent information about him that was obtained for one purpose from being used or made available for other purposes without his consent."[199] In the Charter of Fundamental Rights of the European Union, the purpose specification is included in the right to protection of personal data.[200]

The requirement that personal data may only be used for purposes that are "not incompatible" is somewhat vague. The Article 29 Working Party, an advisory body in which national data protection authorities from Europe cooperate,[201] has discussed the purpose specification in depth. To assess whether a new purpose is compatible with the collection purpose, says the Working Party, all circumstances must be considered. Relevant circumstances include the relation between the original and the new purpose, the collection context, the reasonable expectations of the data subject,[202] the personal data's sensitivity, the risks resulting from the new purpose, and the measures the controller has in place to mitigate risks.[203]

---

197. Nigel Shadbolt, Wendy Hall & Tim Berners-Lee, *The Semantic Web Revisited*, 21 IEEE INTELLIGENT SYSTEMS 96, 98 (2006); *see also* WENDY HALL ET AL., NOMINET TRUST, OPEN DATA AND CHARITIES 16 (2012), http://www.nominettrust.org.uk/sites/default/files/Open%20Data%20and%20Charities.pdf ("Open data, taking inspiration from other ideologies of openness such as open source and open access publishing, articulates the idea that data should be usable by anyone, not just the data owner (or 'data controller' in the language of the Data Protection Act).").
198. See the Purpose Specification Principle from the OECD Guidelines, *supra* Section IV.A.2.
199. RECORDS, COMPUTERS, AND THE RIGHTS OF CITIZENS, *supra* note 148, at xx.
200. E.U. Charter of Fundamental Rights, *supra* note 68, art. 8(2).
201. On the Working Party generally, see Yves Poullet & Serge Gutwirth, *The Contribution of the Article 29 Working Party to the Construction of a Harmonised European Data Protection System: An Illustration of "Reflexive Governance"?*, *in* DÉFIS DU DROIT À LA PROTECTION DE LA VIE PRIVÉE [CHALLENGES OF PRIVACY AND DATA PROTECTION LAW] 570 (María Verónica Perez Asinari & Pablo Palazzi eds., 2008). The Working Party's opinions are not legally binding, but they are influential in Europe. Judges and national Data Protection Authorities often follow the Working Party's interpretation.

According to the Working Party, an example of a new purpose that is incompatible with the original processing purpose is contained in the following hypothetical. A public sector body publishes public servants' contact details on its website, to enable the public to contact them.[204] A re-user wants to merge the public servants' home addresses and phone numbers with the published contact details, to build an interactive map.[205] The re-use is not within the reasonable expectations of the civil servants, making the purpose incompatible and thus not allowed.[206]

### 2. Security and Accountability Principles

The security principle requires appropriate security for personal data. Data controllers must protect data against unauthorized disclosure, access, or other use. When thoughtlessly releasing personal data, a public sector body breaches the security principle. After all, the public sector body would have no control over how the data are used—and neither would the data subjects. The mere fact that data subjects have no control over the use of their data is a subjective privacy harm. Moreover, anybody could access the data, including data brokers and identity thieves.

The accountability principle makes the data controller responsible for complying with the FIPs. The OECD Guidelines define the data controller as the party that "is competent to decide about the contents and use of personal data regardless of whether or not such data are collected, stored, processed or disseminated by that party or by an agent on its 2015]   OPEN DATA, PRIVACY, AND FAIR INFORMATION   2111behalf."[207] A public sector body holding the personal data is usually the data controller. If a re-user obtains personal data from the public sector body, the re-user typically becomes a data controller as well.

---

202. In the United States and the European Union, the "reasonable expectation of privacy" is interpreted differently. The European Court of Human Rights says a "person's reasonable expectations as to privacy is a significant though not necessarily conclusive factor." Perry v. United Kingdom, 2003-IX Eur. Ct. H.R. 141, ¶ 37. On the United States, see SOLOVE & SCHWARTZ, *supra* note 66, at 288–335.
203. *Opinion of the Article 29 Data Protection Working Party on Open Data and Public Sector Information ("PSI") Re-use*, at 20, 1021/00/EN WP 207 (June 5, 2013) [hereinafter *Article 29 Opinion on Open Data and PSI Re-use*]; *see also Opinion of the Article 29 Data Protection Working Party on Purpose Limitation*, 00569/13/EN WP 203 (Apr. 2, 2013).
204. *Article 29 Opinion on Open Data and PSI Re-use*, *supra* note 203, at 20.
205. *Id.*
206. *Id.*

### 3. Data Quality Principle

The data quality principle requires appropriate accuracy, completeness, and relevancy of personal data. One of the aims of the principle is to reduce the risk that organizations base decisions about people on incorrect data. Decisions based on incorrect data can have disastrous effects for a data subject.[208] The data quality principle is relevant to open data. Releasing incorrect personal data could have a detrimental effect.[209] For example, imagine that a website about political campaign financing erroneously includes your name as a donor to a fringe extremist party.

### 4. Collection Limitation and Transparency Principle

The transparency principle, or openness principle, requires transparency regarding data processing, especially towards the data subject.[210] The transparency principle aims to prevent data controllers from abusing information asymmetry.

The transparency principle is prominent in data privacy laws, and can be recognized, for instance, in the proposed U.S. Consumer Privacy Bill of Rights,[211] the E.U. Data Protection Directive,[212] and the proposed E.U. Data Protection Regulation.[213] Some authors suggest that the transparency

---

207. OECD Privacy Guidelines, *supra* note 158, art. 1. The OECD data controller concept is different from the E.U. concept of "data controller." In brief, under E.U. data protection law, the data controller is the party that determines the goals and means for personal data processing. A party that processes personal data on behalf of the controller is the "data processor." Directive 95/46/EC, *supra* note 69, art. 2(d)–(e).

208. *See, e.g.*, Romet v. Netherlands, Eur. Ct. H.R. No. 7094/06 (2012).

209. *See* Scassa, *supra* note 71; Rotenberg, *supra* note 157.

210. To avoid confusion with the open character of open data, we will speak of the "transparency principle" rather than of the "openness principle."

211. CONSUMER DATA PRIVACY IN A NETWORKED WORLD, *supra* note 151, at 47 (discussing the Consumer Privacy Bill of Rights and transparency principle).

212. Directive 95/46/EC, *supra* note 69, arts. 10, 11.

213. *Proposal for a Regulation of the European Parliament and of the Council on the Protection of Individuals with Regard to the Processing of Personal Data and on the Free Movement of Such Data (General Data Protection Regulation)*, art. 5(a), COM (2012) 11 final (Jan. 25, 2012). In December 2015, agreement was reached on the Regulation's text. *See Regulation (EU) No. XXX/2016 of the European Parliament and of the Council on the Protection of Individuals with Regard to the Processing of Personal Data and on the Free Movement of Such Data (General Data Protection Regulation)*, COM (2016) 15039/15 limite (Dec. 15, 2015), http://data.consilium.europa.eu/doc/document/ST-15039-2015-INIT/en/pdf. At the time of writing, the European Parliament and the Council must principle is the most important principle of the FIPs.[214] The transparency principle has old roots. The first principle of the U.S. Department of Health, Education, and Welfare report of 1973 states "[t]here must be no personal-data record-keeping systems whose very existence is secret."[215] The second principle adds that "[t]here must be a way for an individual to find out what information about him is in a record and how it is used."[216]

The collection limitation principle requires that personal data, where appropriate, be collected with the data subject's knowledge or consent. The Article 29 Working Party recommends that a public sector body inform data subjects in advance whether the personal data they provide might be disclosed, for example due to freedom of information laws.[217]

### 5. *Use Limitation and Individual Participation Principle*

The individual participation principle aims to give people some control over the processing of their personal data. For instance, data subjects have the right, under certain circumstances, to rectify their data. The principle illustrates that the privacy as control perspective has influenced the FIPs.[218]

As previously stated, unrestricted re-use of personal data would breach the purpose specification principle—but the use limitation principle seems to offer a way out. The use limitation principle says that personal data should only be used in accordance with the purpose specification principle, except "a) with the consent of the data subject; or b) by the authority of law."[219] Hence, personal data can be used for a new (prima facie incompatible) purpose if the data subject consents to the new use. Indeed, some have suggested that public sector bodies should obtain consent of the relevant individuals before releasing data as open data.[220]

---

still formally adopt the final text. The Regulation is expected to become applicable in 2018.

 214. *See, e.g.*, de Hert & Gutwirth, *supra* note 157; ZUIDERVEEN BORGESIUS, IMPROVING PRIVACY PROTECTION, *supra* note 73, at 99, 106–11.
 215. RECORDS, COMPUTERS, AND THE RIGHTS OF CITIZENS, *supra* note 148, at 41.
 216. *Id.*
 217. *Article 29 Opinion on Open Data and PSI Re-use*, *supra* note 203, at 9. Arvind Narayanan et al. suggest that people should be informed regarding re-identification risks. ARVIND NARAYANAN, JOANNA HUEY & EDWARD W. FELTEN, A PRECAUTIONARY APPROACH TO BIG DATA PRIVACY 1, 16 (2015).
 218. *See* Kirby, *supra* note 163, at 8 (citing Alan Westin as an influence on the OECD Guidelines).
 219. OECD PRIVACY FRAMEWORK, *supra* note 173, ¶ 10 (Use Limitation Principle).
 220. *See, e.g.*, Bart van der Sloot, *On the Fabrication of Sausages, or of Open Government and Private Data* 3 JeDEM 1, 14 (2011).

However, relying on data subject consent for disclosing personal data as open data has some drawbacks. First, people are often in a dependent position vis-à-vis the public sector, and that position may make consent involuntary. Somebody interacting with the public sector might not feel free to withhold consent. Say Alice goes to a city council office for unemployment benefits. Alice really needs money, as she has missed five rent payments, and risks being evicted with her young child. Because she wants to be cooperative, Alice is unlikely to withhold consent to any request by the city council office. Under E.U. data privacy law, consent given under too much pressure is invalid, because consent must be "freely given."[221] For instance, if an employer asks an employee for consent, the consent might not be freely given because of the power imbalance.[222] And according to the European Court of Justice, people applying for passports cannot be deemed to have freely consented to have their fingerprints taken, because people need a passport.[223]

A second problem with data subject consent as a justification for disclosing personal data is that a request for consent can only be meaningful if it specifies a processing purpose.[224] A third problem is that behavioral studies cast doubt on individual consent as a privacy protection measure. For example, on the Internet, people tend to click "I agree" to requests that they see on their screens without knowing what they are agreeing to.[225] Furthermore, it may be impractical for the public sector body to obtain the consent of thousands of individuals. In sum, obtaining data subjects' consent to release personal data is not a general solution to reconcile FIPs and open data policy.

To conclude, from a FIPs perspective, the main problem with open data is that it can be used by anyone, for any purpose, without re-use restrictions. A complete lack of re-use restrictions clashes with the purpose specification principle.

---

221. ELENI KOSTA, CONSENT IN EUROPEAN DATA PROTECTION LAW 256 (2013).
222. *Opinion of the Article 29 Data Protection Working Party on the Definition of Consent*, at 13–14, 01197/11/EN WP187 (July 13, 2011) [hereinafter *Article 29 Opinion on Consent*].
223. C-291/12, Schwarz v. Stadt Bochum, EUR-Lex CELEX LEXIS 0291 ¶ 32 (Oct. 17, 2013) (CJEU).
224. *Article 29 Opinion on Consent*, *supra* note 222, at 9.
225. *See, e.g.*, Acquisti & Grossklags, *supra* note 93; Daniel J. Solove, *Introduction: Privacy Self-Management and the Consent Dilemma*, 126 HARV. L. REV. 1880 (2013); Solon Barocas & Helen Nissenbaum, *Big Data's End Run Around Anonymity and Consent*, *in* PRIVACY, BIG DATA, AND THE PUBLIC GOOD: FRAMEWORKS FOR ENGAGEMENT 44 (Julia Lane et al. eds., 2014); ZUIDERVEEN BORGESIUS, IMPROVING PRIVACY PROTECTION, *supra* note 73.

## V. TYPES OF DATA

Compromises are possible to balance privacy and open data interests. This balancing act may play out differently for different types of data. To help balance the different interests, we distinguish between four data categories, with different levels of privacy risks: (A) raw personal data, (B) pseudonymized data, (C) anonymized data, and (D) non-personal data. We borrow the "raw personal data" category from Tim Davies, and borrow the other three categories from the Article 29 Working Party.[226] We distinguish the four categories to structure the discussion, but the boundaries between them are not clear-cut, as noted in Section V.E.

### A. RAW PERSONAL DATA

With raw personal data, no attempt has been made to mitigate re-identification risks. Examples of raw personal data include names, social security numbers, and personal email addresses. Some open data advocates suggest that open data should never include raw personal data.[227] Indeed, while open data policy is important, releasing raw personal data without any re-use restrictions is usually neither desirable nor legally feasible.

However, in some circumstances raw personal data should be disclosed, because the public interests in disclosure outweigh the privacy interests. For instance, say a public registry of judges reveals positions and jobs judges hold elsewhere, to uphold impartiality of the judiciary. If these data did not identify individual judges, disclosure would not offer sufficient transparency.[228] More generally, people must accept that their privacy diminishes if they take on certain functions in the public sector. For example, it is widely accepted that media can report on politicians, even when politicians might sometimes prefer that certain information remain confidential.[229]

---

226. Tim Davies, *Untangling the Data Debate: Definitions and Implications*, OPENDATAIMPACTS.NET (Mar. 23, 2012), http://www.opendataimpacts.net/2012/03/untangling-the-open-data-debate-definitions-and-implications; *Opinion of the Article 29 Data Protection Working Party on Anonymisation Techniques*, 0829/14/EN WP 216, (Apr. 10, 2014) [hereinafter *Article 29 Opinion on Anonymisation Techniques*]. *See also* HALL ET AL., NOMINET TRUST, *supra* note 197.

227. For instance, Wendy Hall et al. say that raw personal data "should never be directly published as openly licensed and accessible data without explicit consent of the individuals covered in the data." HALL ET AL., NOMINET TRUST, *supra* note 197, at 14. Tim Berners-Lee and Nigel Shadbolt, two authors who promote open data, say that "[i]n the drive to free up data we have always argued that it is essential to respect individual privacy and national security." Tim Berners-Lee & Nigel Shadbolt, *There's Gold to be Mined from All Our Data*, TIMES (London), Dec. 31, 2011.

But the fact that certain raw personal data should be disclosed does not imply that they should be disclosed as open data without re-use restrictions. Even if a law states that certain information must be made public, it does not necessarily follow that such information should be released fully openly. By 1972, some already argued that "the assumptions built into 19th century ideals of public records need revisiting in light of technology."[230] And as Scassa notes, "[i]n many cases, decisions around the public nature of the information were made in an era before the Internet."[231] Hence, personal data that are required to be public by law should not automatically be seen as data that can be released as fully open data.

To illustrate, in many countries court proceedings are mandated to be public. But if court proceedings can only be consulted by traveling to the courthouse and inspecting paper files, the personal information in those files is protected by "practical obscurity."[232] As the U.S. Supreme Court noted in 1989, "there is a vast difference between the public records that might be found after a diligent search of courthouse files, county archives, and local police stations throughout the country and a computerized summary located in a single clearinghouse of information."[233]

In sum, even if the law requires disclosing certain personal information as part of the public record, the public sector body should still assess whether this information should also be made available as open data on the web.[234] As we shall argue in Part VI, the question of whether or not data should be made available as open data is a further, additional question that follows the question of whether the data should be made publicly available at all.

---

228. As Gary T. Marx puts it, sometimes "disclosure norms" trump "privacy norms." Gary T. Marx, *Foreword: Privacy Is Not Quite Like the Weather*, in PRIVACY IMPACT ASSESSMENT v, viii (David Wright & Paul De Hert eds., 2012).
229. *See supra* Section III.C.
230. Chris Jay Hoofnagle, Summary, *Archive of the Meetings of the Secretary's Advisory Committee on Automated Personal Data Systems (SACAPDS)*, BERKELEY LAW (July 15, 2014), https://www.law.berkeley.edu/centers/bclt/research/privacy-at-bclt/archive-of-the-meetings-of-the-secretarys-advisory-committee-on-automated-personal-data-systems-sacapds.
231. Scassa, *supra* note 71, at 403; *see also* Keenan, *supra* note 97, at 1.
232. U.S. Dep't of Justice v. Reporters Comm. for Freedom of the Press, 489 U.S. 749, 762 (1989).
233. *Id.* at 764.

B.   PSEUDONYMIZED DATA

Pseudonymized data are personal data about an individual that are tied to a unique identifier other than a name. For instance, "William Carey Jones" could be referred to as person number "4.417.749." Pseudonymization can be described as follows: "replacing one attribute (typically a unique attribute) in a record by another."[235] Merely substituting names with other unique identifiers is rarely enough to anonymize personal data, or to safeguard privacy.[236]

A well-known example of the limited effect of pseudonymization as an anonymization measure is the 2006 AOL data breach. AOL released pseudonymized data about users of its search engine by replacing the name of each searcher with a number.[237] However, journalists soon found out the real name of the person behind one of the pseudonymous search profiles and published an article entitled *A Face is Exposed for AOL Searcher No. 4417749*.[238] The journalists found the woman behind search profile 4417749 without using sophisticated re-identification techniques.[239] The search queries of user number 4417749 suggested that the searcher was an elderly woman with a dog, from a specific town.[240] When the journalists visited her house, she confirmed that the searches were hers.[241]

The Article 29 Working Party suggests that pseudonymized data are a type of personal data, and are thus within the scope of European data protection law.[242] Some computer scientists have a similar view.[243]

---

234. *See generally* Amanda Conley, Anupam Datta, Helen Nissenbaum & Divya Sharma, *Sustaining Privacy and Open Justice in the Transition to Online Court Records: A Multidisciplinary Inquiry*, 71 MD. L. REV. 772 (2012).
235. *Article 29 Opinion on Anonymisation Techniques*, *supra* note 226, at 20.
236. NARAYANAN, HUEY & FELTEN, *supra* note 217, at 2; PRESIDENT'S COUNCIL OF ADVISORS ON SCI. & TECH., EXEC. OFFICE OF THE PRESIDENT, BIG DATA AND PRIVACY: A TECHNOLOGICAL PERSPECTIVE, at 38–39 (2014), http://www.whitehouse.gov/sites/default/files/microsites/ostp/PCAST/pcast_big_data_and_privacy_-_may_2014.pdf.
237. Michael Barbaro & Tom Zeller Jr., *A Face is Exposed for AOL Searcher No. 4417749*, N.Y. TIMES, Aug. 9, 2006.
238. *Id.*
239. *Id.*
240. *Id.*
241. *Id.*
242. *Article 29 Opinion on Anonymisation Techniques*, *supra* note 226, at 10.
243. NARAYANAN, HUEY & FELTEN, *supra* note 217.

However, the Working Party's view has also been criticized for making the scope of personal data too broad.[244]

While pseudonymizing personal data rarely, if ever, makes people non-identifiable, pseudonymization can help to protect privacy interests, by making it a bit harder to recognize people by name.[245] For instance, Conley et al. suggest that pseudonymization can help to mitigate privacy concerns when court cases are published online.[246] If people's names are changed to "[party 1]" and "[party 2]" in judgments, it would be impossible to search within court records on the basis of a person's name. Pseudonymization also reduces the chance that somebody who looks at the data will recognize a person by name. However, it might still be possible to recognize people based on the facts of a case discussed in the judgment. Nevertheless, pseudonymization adds a thin layer of practical obscurity.[247]

Different countries have different traditions. In the Netherlands, many court decisions are published online on a centralized website.[248] But if the litigating parties are individuals, their names are changed to neutral phrases such as "plaintiff" and "defendant."[249] In other countries, litigants' names are often included in court documents, even when published online.[250]

---

244. *See, e.g.*, Khaled El Emam & Cecilia Álvarez, *A Critical Appraisal of the Article 29 Working Party Opinion 05/2014 on Data Anonymization Techniques*, 5 INT'L DATA PRIVACY L. 73 (2015).

245. NARAYANAN, HUEY & FELTEN, *supra* note 217, at 2; PRESIDENT'S COUNCIL OF ADVISORS ON SCI. & TECH., *supra* note 236, at 38–39.

246. Conley, Datta, Nissenbaum & Sharma, *supra* note 234, at 842.

247. *See id.*

248. *See generally* Laurens Mommers, *Access to Law in Europe*, in INNOVATING GOVERNMENT 383 (Simone van der Hof & Marga M. Groothuis eds., 2011); LEONIE VAN LENT, EXTERNE OPENBAARHEID IN HET STRAFPROCES (2008).

249. These are our translations. *Anonimiseringsrichtlijnen*, RECHTSPRAAK, https://www.rechtspraak.nl/Uitspraken-en-nieuws/Uitspraken/Paginas/Anonimiserings richtlijnen.aspx (last visited July 2, 2015). Not all personal data are obfuscated; for example, the attorneys for a case are mentioned by name. The Spanish system is similar to the Dutch one. *See* James B. Jacobs & Elena Laurrauri, *Are Criminal Convictions a Public Matter? The USA and Spain*, 14 PUNISHMENT & SOC'Y 3 (2012) (with further references to literature on other countries).

250. About the United States, see Nancy S. Marder, *From "Practical Obscurity" to Web Disclosure: A New Understanding of Public Information*, 59 SYRACUSE L. REV. 441, 444–47 (2009); Conley, Datta, Nissenbaum & Sharma, *supra* note 234. For an overview of how to obtain criminal records in fifty-four countries, see KPMG, DISCLOSURE OF CRIMINAL RECORDS IN OVERSEAS JURISDICTIONS (2009), http://www.cpni.gov.uk/ documents/publications/2009/2009-criminal_records_disclosure_intro_and_exe_summary _march09.pdf?epslanguage=en-gb.

In sum, pseudonymization can help to reduce privacy risks—it is a useful but not sufficient security measure. Because pseudonymizing data is not enough to anonymize data, pseudonymous data must generally be treated as personal data.

C. ANONYMIZED DATA

Anonymized data are ex-personal data that are rendered anonymous in such a way that data subjects are no longer identifiable. Aggregated data are typically anonymous. For instance, the information that "112,580 people live in Berkeley," without additional information, does not identify an individual. Anonymization can be defined as "a technique applied to personal data in order to achieve irreversible de-identification."[251] Anonymized data are outside the scope of the FIPs, as the FIPs only apply to personal data.

The fact that personal data can be aggregated and thereby anonymized seems an appropriate way to strike a balance between privacy interests and open data interests.[252] For instance, statistics can often be disclosed as open data, as long as they are anonymized and aggregated.[253] To illustrate, a crime map could say that on a certain day, "between one and ten burglaries took place on or near Bancroft Way," rather than "one burglary took place at 11 Bancroft Way." Traffic data could say that "between one and ten cars drove on Bancroft Way between April 15 and April 20, 2015," rather than "one car drove on Bancroft Way on April 19, 2015 at 12:24 A.M."

But two caveats are in order. First, anonymizing data does not guarantee privacy and fairness.[254] For instance, the Dutch public reacted angrily when the police used aggregated information derived from data gathered by TomTom, a vendor of car navigation systems.[255] The police used the data to choose the best spots to install speeding cameras.[256] The Dutch Data Protection Authority examined whether TomTom's practices complied with E.U. data privacy law, and did not find major problems.[257] The data obtained by the police were properly anonymized through aggregation, and thus outside the scope of the FIPs.[258]

---

251. *Article 29 Data Protection Working Party on Anonymisation Techniques*, *supra* note 226, at 7.
252. *See Article 29 Opinion on Open Data and PSI Re-use*, *supra* note 203, at 12; HALL ET AL., NOMINET TRUST, *supra* note 197, at 45; *see also* OFFICE OF THE AUSTRALIAN INFO COMM'R, INFORMATION POLICY AGENCY RESOURCE 1: DE-IDENTIFICATION OF DATA AND INFORMATION, (2014), http://www.oaic.gov.au/images/documents/information-policy/information-policy-resources/information-policy-agency-resources/information_policy_agency_resource_1.pdf.
253. *See* Francesco Molinari & Jesse Marsh, *Does Privacy Have to Do with Open Data? Some Preliminary Reflections—and Answers*, *in* CEDEM13 CONFERENCE FOR E-DEMOCRACY AND OPEN GOVERNMENT 303, 311 (Peter Parycek & Noella Edelmann eds., 2013).
254. NARAYANAN, HUEY & FELTEN, *supra* note 217, at 3.
255. Charles Arthur, *TomTom Satnav Data Used to Set Police Speed Traps*, GUARDIAN, Apr. 28, 2011.

The TomTom example illustrates a broader problem: the FIPs apply to personal data—and only to personal data. But people can be treated unfairly, or feel like they are being treated unfairly, on the basis of information that is *based* on personal data concerning them, but that is not personal data anymore.[259] Moreover, as the aggregated information is outside the scope of the FIPs, the data subject rights that follow from the FIPs, such as access and correction rights, no longer apply. As Seda Gürses notes, anonymization can "disempower" the individual.[260] The FIPs and most data privacy laws around the world have this problem in common.[261] We will not attempt to solve the problem here. But we do note that sometimes a public sector body may want to decide not to release anonymized information, even if the information is outside the of FIPs' scope.

A second caveat is that anonymized data are often less interesting for re-users than raw personal data or pseudonymous data. As Bendert Zevenbergen et al. put it:

> The utility and privacy of data are generally directly and inversely related. For many datasets, it has proven difficult—if not impossible—to increase data subjects' privacy without concurrently decreasing the overall utility of the dataset. Small privacy gains are generally achieved by far-reaching decreases in

---

    256. *Id.*
    257. Press Release, *Following Report by Dutch DPA, TomTom Provides User with Better Information*, COLLEGE BESCHERMING PERSOONSGEGEVENS (Jan. 12, 2012), https://cbpweb.nl/en/news/following-report-dutch-dpa-tomtom-provides-user-better-information.
    258. *See id.*; *see also* Harold Goddijn, *This is What We Really do with Your Data*, TOMTOM.COM, http://www.tomtom.com/page/facts (last visited June 23, 2015).
    259. *See generally* Lyon, *Surveillance as Social Sorting*, *supra* note 102.
    260. Seda Gürses, *The Spectre of Anonymity*, *in* SNIFF, SCRAPE, CRAWL . . . : ON PRIVACY, SURVEILLANCE AND OUR SHADOWY DATA-DOUBLE 52 (2012).
    261. *See generally* PROFILING THE EUROPEAN CITIZEN: CROSS-DISCIPLINARY PERSPECTIVES (Mireille Hildebrandt & Serge Gutwirth eds., 2008); Barocas & Nissenbaum, *supra* note 225; Joris Van Hoboken & Frederik Zuiderveen Borgesius, Scoping Electronic Communication Privacy Rules: Data, Services or Values (2015).

> data utility. A small increase in data utility often requires much more personal information to be revealed.[262]

In sum, anonymized data can—in theory—safely be disclosed as open data, without re-use restrictions. However, in practice, irreversible anonymization is exceedingly difficult, and perhaps impossible.

### D. NON-PERSONAL DATA

A fourth type of data is non-personal data. Many datasets do not contain, and have never contained, personal data. Examples include datasets regarding public transport times, weather conditions, sea tides, road maps, public sector budgets, and environmental pollution.[263] Such datasets have little to do with information about individuals, and do not fall under the purview of the FIPs.

The FIPs do not hinder releasing datasets with non-personal data. Hence, strict compliance with the FIPs does not necessarily interfere with releasing public sector information. Some suggest that "[m]ost open datasets have nothing personal to be protected in them (e.g.: digital maps, public budgets, air pollution measurements etc.)."[264]

But even for non-personal data, there are caveats. First, sometimes there may be non-privacy related arguments against releasing data. For instance, some information may have to remain confidential because of state security, such as information regarding critical infrastructure locations.[265] Second, as discussed in the next section, a dataset with non-personal data may, on closer inspection, include information about an individual.

In sum, we distinguish between four data categories with different risk levels: raw personal data, pseudonymous data, anonymized data, and non-personal data. However, the categories cannot be neatly distinguished in practice, as discussed next.

---

262. BENDERT ZEVENBERGEN, IAN BROWN, JOSS WRIGHT & DAVID ERDOS, OXFORD INTERNET INST., ETHICAL PRIVACY GUIDELINES FOR MOBILE CONNECTIVITY MEASUREMENTS 11 (2013), http://www.oii.ox.ac.uk/research/Ethical_Privacy_Guidelines_for_Mobile_Connectivity_Measurements.pdf. Along similar lines, see Paul Ohm, *Broken Promises of Privacy: Responding to the Surprising Failure of Anonymization*, 57 UCLA L. REV. 1701 (2010). Slightly more optimistic is Felix T. Wu, *Defining Privacy and Utility in Data Sets*, 84 U. COLO. L. REV. 1117 (2013).
263. Molinari & Marsh, *supra* note 253, at 311.
264. *Id.*; *see also* NARAYANAN, HUEY & FELTEN, *supra* note 217, at 21.
265. *See* Conley, Datta, Nissenbaum & Sharma, *supra* note 234, at 827.

E.   FUZZY BOUNDARIES

The borders between the four data categories are fuzzy. While many data privacy laws make a distinction between personal data and anonymized data, computer science suggests that the distinction is a matter of degree rather than kind.[266] Irreversible anonymization is difficult—perhaps impossible.[267] Apart from that, it is possible to distinguish sub-categories within the four categories. For instance, Zevenbergen et al. distinguish between three types of purportedly anonymized data, with different levels of re-identification risk.[268] And it is debatable whether data about an individual tied to his or her social security number should be seen as raw personal data or as pseudonymized data.

Whether data are sufficiently anonymized is difficult to assess in advance. This is especially so, as more datasets may become available that enable "jigsaw identification."[269] The more data public sector bodies release, the higher the potential for combining data and thus creating information that can identify people.[270] The Obama administration recognizes this, and urges departments and agencies to perform a risk analysis.[271]

Even purportedly non-personal data can provide information about an individual. For example, a dataset with local air pollution levels contains non-personal data. However, if the dataset says that zip code 94720 is the most polluted, and the only business in that zip code is a one-man business, the pollution level in that zip code can say something about the business owner—namely that he or she is likely polluting. We do not suggest that privacy should enable business owners to escape responsibility for polluting. We merely want to illustrate that even datasets with non-

---

266. *See, e.g.*, Arvind Narayanan & Vitaly Shmatikov, *Myths and Fallacies of "Personally Identifiable Information*,*"* 53 COMM. ACM 24 (2010); Ohm, *supra* note 262; Matthijs R. Koot, Measuring and Predicting Anonymity (2012) (Ph.D thesis, University of Amsterdam), https://cyberwar.nl/d/PhD-thesis_Measuring-and-Predicting-Anonymity_2012.pdf.
267. Narayanan & Shmatikov, *supra* note 266, at 26.
268. ZEVENBERGEN, BROWN, WRIGHT & ERDOS, *supra* note 262, at 22.
269. NARAYANAN, HUEY & FELTEN, *supra* note 217, at 5–7. The phrase "jigsaw identification" is from Kieron O'Hara's book. O'HARA, *supra* note 74, at 40.
270. NARAYANAN, HUEY & FELTEN, *supra* note 217, at 5–7; *see also* U.S. GOV'T ACCOUNTABILITY OFFICE, *supra* note 79, at 107.
271. There, the risk is called the "mosaic effect." OMB MEMORANDUM M-13-13, OPEN DATA POLICY, *supra* note 25, at 9–10.

personal data can provide information about an individual, for instance after linking datasets.[272]

In conclusion, we distinguish between four data categories with different risks levels: raw personal data, pseudonymous data, anonymized data, and non-personal data. The next section shows that open data should not be considered the only route when arguments for disclosure outweigh privacy interests. Options other than releasing data as open data are also available, such as disclosing data with access or re-use restrictions.

## VI.    TYPES OF DISCLOSURE

A maximalist approach to publishing public sector information as open data might imply that a public sector body should not impose any conditions on accessing or re-using public sector information. But a more moderate view is that public sector bodies should be allowed to impose conditions for access and re-use, if this is required to protect privacy interests.

We distinguish between three types of disclosure with different degrees of openness: (A) restricted access, (B) restricted use, and (C) open data. Restrictions on access and restrictions on re-use can be combined. Some forms of access and re-use restrictions do not comply with certain definitions of "open" data.[273] But sometimes disclosing data with restrictions is better than not disclosing at all.[274]

### A.    DISCLOSURE WITH ACCESS RESTRICTIONS

The first way to balance privacy and open data policy is by restricting access. To achieve a particular objective that underpins open data, it might not be necessary to allow everyone access, or to allow access to the raw data held by a public sector body. Data can be disclosed to particular groups for particular purposes, rather than to anybody for any purpose. Completely blocking data release on the one hand, and releasing data as open data on the other hand, can be seen as two extremes on a continuum. Disclosing data with restrictions is in between those two extremes.

---

    272. For example, a health insurance company might use the data to calculate the health risks of the pollution, and might charge some people higher prices for coverage.
    273. For example, if data are made available for non-commercial uses only, this runs counter to the open data principles set out in Part II ("open" implies that data can be used for any purpose). The same is true if only certain types of users are given access ("open" implies that data can be used by anyone). *See supra* Section II.A.
    274. Scassa arrives at a similar conclusion about balancing privacy and public access. "As the experience of courts and tribunals shows, it may sometimes be necessary to place limits on the digital disclosure of some of the information in a 'public' record in order to achieve this balance." Scassa, *supra* note 71, at 404.

There are various ways to disclose data, which bring different risk levels. For example, Zevenbergen et al. distinguish open data from "restricted" disclosure, "managed access," "interactive methods," and "hybrid" methods.[275]

With "restricted" disclosure, data are only disclosed "to persons or organisations on request, refusing dissemination when the level of risk is considered too high."[276] Zevenbergen et al. suggest, for instance, that it is riskier to disclose data to a company than to academic researchers.[277] One problem with this type of disclosure is that it is hard to monitor what a receiving party does with the data.

With managed access, "[t]hird parties can query the dataset and conduct statistical (or other) analysis. Such an approach allows the researcher to ascertain exactly who accesses the datasets, while maintaining control over its dissemination."[278] For instance, researchers might have to visit the offices of the public sector body to inspect data.[279]

An example of an interactive method is "differential privacy."[280] As Zevenbergen et al. explain:

> Differential Privacy . . . only gives statistical answers to queries about an underlying dataset. To protect privacy even further, a certain amount of noise is added to the disclosed statistical data. In principle, differential privacy offers a lower risk for privacy, but there are certain limitations to this approach that need to be understood. For example, the uncertainty related by the addition of noise to the data can be exhausted, which means the dissemination must then stop.[281]

---

    275. ZEVENBERGEN, BROWN, WRIGHT & ERDOS, *supra* note 262, at 28–29.
    276. *Id.* at 28.
    277. *Id.* at 14–16. Similarly, Narayanan et al. say that restricted access "is a good solution" to enable scientific research without releasing data as fully open data. NARAYANAN, HUEY & FELTEN, *supra* note 217, at 20.
    278. ZEVENBERGEN, BROWN, WRIGHT & ERDOS, *supra* note 262, at 29.
    279. *See id.* at 15.
    280. Cynthia Dwork, *Differential Privacy*, *in* ENCYCLOPEDIA OF CRYPTOGRAPHY AND SECURITY 338, 338–40 (Henk C.A. van Tilborg & Sushil Jajodia eds., 2011).
    281. ZEVENBERGEN, BROWN, WRIGHT & ERDOS, *supra* note 262, at 29 (emphasis omitted).

Hybrid approaches are also possible. For instance, parts of a dataset could be disclosed publicly, while other parts of the set could be kept confidential, or could be disclosed with strict access restrictions.[282]

In sum, sometimes a compromise between openness and privacy can be found by releasing data with access restrictions. Apart from access restrictions, it is also possible to restrict re-use, as discussed next.

### B. DISCLOSURE WITH RE-USE RESTRICTIONS

Another way to strike a balance between privacy and open data policy is by applying restrictions on re-use of the disclosed data.[283] For instance, re-use restrictions can come in the form of licenses.[284] The license could require re-users not to re-identify data. Such measures have been used in practice. For example, on the website of the U.S. Healthcare Cost and Utilization Project, data users can purchase data sets—but if they purchase a dataset, they must sign an agreement that "expressly prohibits any attempt to identify individuals."[285]

The Article 29 Working Party suggests that the license should "prohibit license-holders from using the data to take any measure or decision with regard to the individuals concerned."[286] The license should also require "the license-holder to notify the licensor in case it is detected that individuals can be or have been re-identified."[287] As proper anonymization is difficult, in some situations, anonymized datasets should only be released under a license regime, rather than as fully open data. The higher the risk of de-anonymization, the more reason to tie a license to a dataset.

Access and re-use restrictions can also be combined. For instance, researchers could be required to visit the office of a public sector body to inspect a dataset: an access restriction. But at the same time, the researchers could be required to not try to re-identify people in the dataset: a re-use restriction.

---

282. *Id.*
283. Solove makes a similar distinction between "access restrictions" and "use restrictions." Solove, *supra* note 7, at 1169–70.
284. *Article 29 Opinion on Open Data and PSI Re-use*, *supra* note 203, at 25–26; *see also* NARAYANAN, HUEY & FELTEN, *supra* note 217, at 18. An issue that falls outside the scope of this paper is the legal basis for such licenses. In some countries, the public sector might have a type of intellectual property right on the dataset; in other countries the public sector body could invoke general contract law to impose a license on the dataset.
285. SID/SASD/SEDD Application Kit, HEALTHCARE COST & UTILIZATION PROJECT 24 (Sept. 16, 2015), http://www.hcup-us.ahrq.gov/db/state/SIDSASDSEDD_Final.pdf.
286. *Article 29 Opinion on Open Data and PSI Re-use*, *supra* note 203, at 25.
287. *Id.*

C.    DISCLOSURE AS OPEN DATA

The third access type is releasing data as fully open data: with no access or re-use restrictions. For instance, perhaps some personal data included in lobbying or company registers should be released as open data. Restricting access or re-use might make it too difficult to analyze the influence of lobbyists or to hold companies accountable.[288]

In conclusion, sometimes a balance can be struck between open data goals and privacy by disclosing data with access or re-use restrictions, rather than as fully open data. Hence, a public sector body must first assess whether a dataset should be disclosed at all. If it is decided that data should be disclosed, the next question is whether the data should be released with access or re-use restrictions, or as fully open data.

## VII.  A CIRCUMSTANCE CATALOGUE TO INFORM DISCLOSURE DECISIONS

The above suggests that public sector bodies should decide on a case-by-case basis whether, and under which conditions, a dataset should be disclosed.[289] Narayanan et al. note that "[e]ach dataset has its own risk-benefit tradeoff, in which the expected damage done by leaked information must be weighed against the expected benefit from improved analysis."[290] The researchers add that "[b]oth assessments are complicated by the unpredictable effects of combining the dataset with others, which may escalate both the losses and the gains."[291]

There is not one clear-cut rule to decide whether datasets including or based on personal data should be disclosed. The lack of a hard-and-fast rule is not surprising. As discussed *supra* in Part III, the problem of balancing privacy and open data interests can be seen as a modern version of the problem of balancing privacy and public sector transparency.

---

288. *See, e.g.*, Jonathan Gray & Tim Davies, Fighting Phantom Firms in the UK: From Opening Up Datasets to Reshaping Data Infrastructures? (May 27, 2015) (Working paper presented at the Open Data Research Symposium, 3rd International Open Government Data Conference, Ottawa), http://ssrn.com/abstract=2610937; TRANSPARENCY INTERNATIONAL, HOW OPEN DATA CAN HELP TACKLE CORRUPTION (2015), http://www.transparency.org.uk/publications/how-open-data-can-help-tackle-corruption-policy-paper.

289. Many authors arrive at that conclusion. *See, e.g.*, Katleen Janssen & Sara Hugelier, *Open Data: A New Battle in an Old War Between Access and Privacy?*, in DIGITAL ENLIGHTENMENT YEARBOOK 2013, at 190, 199 (Mireille Hildebrant et al. eds., 2013).

290. NARAYANAN, HUEY & FELTEN, *supra* note 217, at 12.

291. *Id.*; *see also id.* at 13, 15.

The objectives behind open data policies and corresponding public interests involved merit closer scrutiny; this allows for differentiation that is necessary for balancing the interests involved.[292] The general FIPs guidance suggesting a balance between privacy and other interests is not detailed enough in the case of open data. We propose that a circumstance catalogue can help to decide whether and how to release data.[293] The circumstance catalogue lists circumstances, or factors, that should be considered when assessing whether, and under which conditions, a dataset should be released, as well as different options for how it should be released. We provide a list as a starting point for a debate—the list is not meant to be exhaustive or final. The circumstance catalogue can be extended, for instance, by taking inspiration from case law, freedom of information law, and guidelines regarding open data and privacy.

We mention some rules of thumb regarding re-identification risks and releasing data. One rule of thumb is that raw personal data should generally not be released as fully open data, unless there is a compelling public interest argument for choosing this route for disclosure over other available options.[294] We argue that pseudonymous data must generally be treated as a type of personal data, rather than as anonymous data. On the other hand, non-personal data can generally be released as open data. For purportedly anonymized data, it is more complicated. As stated previously, irreversible anonymization is difficult, and perhaps impossible to achieve.[295] Therefore, in some cases anonymized data should not be released as fully open data.

## A. WEIGHT OF THE GOALS PURSUED

The goals pursued by disclosing data are relevant. The consideration is not only what the (theoretical) aim of the public body is. An assessment might also be made of the most likely uses of the data by other public bodies, the private sector, and citizens. True, this runs counter to the idea behind open data that serendipitous re-use is positive, and that it is impossible for the government to predict potential uses.[296] But it is naïve to assume that uses will all be benevolent.

---

292. *See* Scassa, *supra* note 71, at 405.
293. For a similar approach, balancing interests in access to court records against other considerations, see Conley, Datta, Nissenbaum & Sharma, *supra* note 234, at 797–98.
294. Narayanan et al. reach a similar conclusion. NARAYANAN, HUEY & FELTEN, *supra* note 217, at 15 ("[I]t almost never will be the case that an unlimited release of a dataset to the entire public will be the optimal choice.").
295. *See supra* Section V.E. *See also* Narayanan & Shmatikov, *supra* note 266, at 26.

What is the primary goal pursued with releasing data and how important is releasing this type of information, in this form, to achieving that goal? Could the objective be adequately addressed by disclosing information in a less privacy-sensitive form? Is it likely that the data will be used primarily by the press or similar public watchdogs, or are the data primarily interesting for commercial purposes?[297] The more relevant data are to key aspects of democratic participation, the stronger the case for release as open data. As Daniel Solove notes, when deciding whether to release personal data, political transparency has more weight than pure commercial interests of re-users:

> Access should be granted for uses furthering traditional functions of transparency such as the watchdog function; access should be denied for commercial solicitation uses because such uses do not adequately serve the functions of transparency. Rather, such uses make public records a cheap marketing tool, resulting in the further spread of personal information, which is often resold among marketers.[298]

Furthermore, not all uses of public sector information are equal before the law. Additionally, the national legal system makes a difference. For instance, the strength of rights to access to information and the discretionary space for public authorities differ from country to country. For example, in the United States, the First Amendment influences decisions regarding data disclosure.[299] In Europe, access to information to foster political transparency also has backing in human rights treaties.[300] But in Europe, legal privacy and data protection rights have more relative weight than in the United States.[301] An important factor in this respect is the role of people whose data are considered for release. Do the data concern somebody who holds a public function or a powerful position? What is the level of responsibility of the person? To what extent is the information needed in open, machine-readable form in order to facilitate democratic accountability? The higher the level of responsibility of the person, the more likely it is that transparency trumps privacy interests.

---

296. For example, the G8 Open Data Charter contains the pledge of governments to ensure "that the data are available to the widest range of users for the widest range of purposes." G8 OPEN DATA CHARTER, *supra* note 1. Assessing the market for public sector information based products and services in the U.K, Deloitte concludes that "it is hard to foresee specifically where innovation might take place in the U.K. Often innovation takes place in areas which are hard to predict." DELOITTE, MARKET ASSESSMENT OF PUBLIC SECTOR INFORMATION, *supra* note 38, at 41.
297. The U.S. Supreme Court noted that different purposes have different weights in the context of inspecting and copying judicial records. Nixon v. Warner Commc'ns, Inc., 435 U.S. 589 (1978). In the FOIA context, the Supreme Court arrived at a similar conclusion. U.S. Dep't of Justice v. Reporters Comm. for Freedom of the Press, 489 U.S. 749, 773 (1989).
298. Solove, *supra* note 7, at 1192.
299. *See id.* at 1200–06; *see also* Sorrell v. IMS Health Inc., 131 S. Ct. 2653 (2011).
300. Mireille van Eechoud & Katleen Janssen, *supra* note 117, at 483–88.

While access to information to foster democratic transparency has backing in constitutions and human rights documents, the legal backing of releasing information for business opportunities or for improving public sector efficiency is less evident.[302] If there is a good case for sharing data within the public sector because this contributes to efficient government, governments should regulate such sharing with specific laws that contain appropriate safeguards. Cost and efficiency savings in and of themselves may not outweigh the protection of individual privacy unless there are other overriding concerns about, for example, public accountability, corruption, or the exercise of democratic oversight.

Apart from the difference in national legal systems, the weight of the goal also depends on the national situation. For example, in a country where there are many problems with corruption by state officials, disclosing detailed wealth records of public functionaries makes more sense than in a country with virtually no corruption. And in some countries there may be more widespread acceptance of the public disclosure of salaries.[303]

B. WEIGHT OF THE PRIVACY INTERESTS

Arguments against releasing data, or against releasing data without restrictions, include the following: there are considerable risks associated with releasing the data; the potential harm is serious, rather than a minor inconvenience; the privacy of many people (not a few) is at risk; and the privacy threat is immediate rather than remote. For instance, a theoretical privacy infringement has less weight than would a clear danger. A clear privacy danger might occur, for example, with a dataset containing names of people with HIV. People could be discriminated against if it becomes publicly known that they have HIV.

---

301. *See generally* Kranenborg, *supra* note 139.
302. The Charter of Fundamental Rights of the European Union does recognize the right to do business. E.U. Charter of Fundamental Rights, *supra* note 68, art. 16.
303. For instance, in Finland, the tax authorities disclose the income of people whose income exceeds certain thresholds. *See* Case C-73/07, Tietosuojavaltuutettu v Satakunnan Markkinapörssi Oy, 2008 E.C.R. I-9831.

The nature of the harm also matters: for example, if the data relate to people fulfilling public functions and concern professional conduct, a risk of reputational harm is unlikely to be of concern (unless there is doubt about the accuracy of the data). That would be the case with disclosing expenses claims. If disclosure leads to a security risk, e.g., disclosing an itinerary or detailed information about a politician's movements, the case is different.

Expectations of privacy can also be a factor. How were the data collected? If there was a promise or understanding of confidentiality, the case is different than if people have volunteered data after they were warned of future possible disclosures. Because of asymmetry in information relationships between public authorities and citizens, it cannot be readily assumed that data was truly volunteered.

In conclusion, a case-by-case analysis is required when deciding whether to release data, and whether the data can be disclosed as fully open data, or whether access or use should be restricted. We proposed a starting point for a circumstance catalogue that would help to assist in decisions about data disclosure.

## VIII.  CONCLUSION

Open data are held to contribute to a wide variety of social and political goals—including strengthening transparency, public participation, and democratic accountability; promoting economic growth and innovation; and enabling greater public sector efficiency and cost savings. But releasing datasets as open data may threaten privacy, for instance if they contain personal or re-identifiable data. Potential privacy problems include chilling effects on people communicating with the public sector, a lack of individual control over personal information, and discriminatory practices enabled by the released data.

Can privacy and related interests be respected, while not unduly hampering open data benefits? The Fair Information Principles (FIPs), as expressed in the OECD Privacy Guidelines, provide a framework to balance privacy and other interests. From a FIPs perspective, the main problem with open data is that it can be used by anyone for any purpose. A complete lack of re-use restrictions would clash with the purpose specification principle of the FIPs. It follows from the purpose specification principle that personal data should only be collected for a purpose that is specified in advance, and that those data should not be used for incompatible purposes.

Compromises are possible to balance privacy and open data interests. We distinguish between four data categories with different risk levels: raw personal data, pseudonymous data, anonymized data, and non-personal data. With raw personal, no attempt has been made to make identification harder. Pseudonymous data are data for which the individual's name is changed to another unique identifier. Anonymized data are ex-personal data; people cannot be re-identified in the dataset. Non-personal data, such as data about weather conditions or public transport times, never contain personal data.

Non-personal data can generally be released without restrictions as fully open data. As a rule of thumb, raw personal data should not be released as fully open data. Pseudonymous data must generally be treated as a type of personal data—not as anonymous data. Anonymized data is more complicated. Anonymized data can, in theory, be disclosed as open data, without re-use restrictions. However, irreversible anonymization is exceedingly difficult, and perhaps impossible. And even in aggregated and purportedly anonymized data, individuals can sometimes be re-identified. Therefore, some purportedly anonymized datasets should only be disclosed with access and re-use restrictions.

Sometimes, a compromise can be found by releasing anonymized data with access and re-use restrictions. Restricting openness can be done in various ways. For instance, the public sector body could attach a license to the data, requiring the re-user to only use certain data for a certain purpose (say medical research) and to promise not to re-identify the data. Other limitations on openness can also be envisaged. For instance, if a research interest is important, but the personal data are sensitive, researchers could be required to visit the lab where the data are held.

Hence, a case-by-case analysis is required when deciding whether to release data, and whether the data can be disclosed as fully open data, or whether access or use should be restricted. To assist in decisions about data disclosure, a circumstance catalogue may be of help: a list of circumstances to consider when deciding about releasing data. For instance: what is the goal pursued by releasing the data? Is there another way to pursue that goal? What are the risks involved with releasing the data? Are the privacy-related risks negligible or probable? If the risk materializes, what is the harm that results? Is the privacy of a few or of millions of people at stake?

In conclusion, in many instances public sector datasets that contain, or are based on, personal data should not be released as fully open data. When arguments for disclosure do outweigh privacy interests, open data should not be considered the only route. Other options might include disclosing information with access or re-use restrictions.